\documentclass{aastex62}

\usepackage{graphicx}

\usepackage{natbib}

\accepted{\today}
\submitjournal{ApJ}

\shorttitle{What sets the characteristic mass of stars}
\shortauthors{Hennebelle et al.}

\newcommand{\sound}{{\mathrm{s}}}

\newcommand{\ad}{{\mathrm{ad}}}

\newcommand{\p}{_\mathrm{p}}
\newcommand{\e}{_\mathrm{e}}
\def\mathbi#1{\textbf{\em #1}}
\def\msol{{\rm M}_\odot}

\begin{document}

   \title{How  first hydrostatic cores, tidal forces and gravo-turbulent fluctuations
 set the characteristic mass of stars}

\correspondingauthor{Patrick Hennebelle}
\email{patrick.hennebelle@cea.fr}

\author{Patrick Hennebelle}
\affiliation{Laboratoire AIM, Paris-Saclay \\
 CEA/IRFU/SAp -- CNRS -- \\
      Universit\'e Paris Diderot, 91191 Gif-sur-Yvette Cedex, France}

\author{Yueh-Ning Lee}
\affiliation{
       Institut de Physique du Globe de Paris \\
 Sorbonne Paris Cit\'e, Universrt\'e  Paris Diderot \\
 UMR 7154 CNRS, F-75005 Paris, France}

\author{Gilles Chabrier}
\affiliation{
      \'Ecole normale sup\'erieure de Lyon \\
 CRAL, UMR CNRS 5574, 69364, Lyon Cedex 07, France}
\affiliation{
      School of Physics \\
 University of Exeter, Exeter, EX4 4QL, UK}


\begin{abstract}
The stellar initial mass function (IMF) is  playing a critical role in
		the history of our universe. 
We propose a theory and that is based solely on local processes, namely the dust 
opacity limit, the tidal forces and the properties of the collapsing gas envelope. The idea is that the final 
mass of the central object is determined by the location of the nearest fragments, which accrete 
the gas located further away, preventing it to fall onto the central object.
To estimate the relevant statistics in the neighbourhood of an accreting protostar, we 
perform high resolution numerical simulations. We also use these simulations 
to further test the idea that fragmentation in the vicinity of an existing protostar is 
determinant in setting the peak of the stellar spectrum. 
We develop an analytical model, which is based on a statistical counting of 
the turbulent density fluctuations, generated during the collapse, that are 
 at least equal to the mass of the first hydrostatic  core, 
and sufficiently important to supersede tidal and pressure forces to be self-gravitating. 
The analytical mass function  presents a peak located at roughly 
10 times  the mass of the first hydrostatic core  in good 
agreement with the numerical simulations. 
 Since the physical processes involved are all local, i.e. occurs at scales of a few 100 AU or below, 
 and do not depend on the gas distribution at large scale and 
global properties such as the mean Jeans mass, 
the mass spectrum is expected to be relatively universal.  
\end{abstract}




\section{Introduction}

The question of the origin of the mass distribution of stars, the initial 
mass function (IMF), is a long standing issue 
\citep[e.g.,][]{salpeter55,kroupa2001,chabrier2003,bastian2010,offner2014} and several  
authors have attempted to provide explanations either using 
analytical modeling of the gas fragmentation \citep{inutsuka2001,padoan1997,HC08},
numerical simulations of a fragmenting cloud \citep{Girichidis11,Bonnell11,BallesterosParedes15}
or statistical description of accretion \citep{basu2004,basu2015,maschberger2014}.
In general, these models have been resonably succesful in reproducing the high mass tail 
of the IMF and obtained powerlaw mass spectra with a slope close to the Salpeter value. 
What is truly intriguing however
is the lack of variations of the IMF given the large amount of 
variations of the star forming environments and particularly the peak 
of the distribution, around 0.3 $M_\odot$, which constitutes a characteristic mass
for the stars. Traditional explanations based on turbulence and 
 Jeans mass \citep{padoan1997,HC08,hopkins2012}, gas cooling and variation of the equation of state
\citep{jappsen2005} or radiative feedback 
\citep{Bate09b,krumholz2016,guszejnov2016}
are all facing a  difficulty to explain the lack of variations of the peak position and require 
the scale dependence of some quantities, such as for instance the Mach number and the density,
or the radiative feedback and the opacities,  to cancel out.

Recently \citet{leeh2018a,leeh18b} (hereafter paper I and II) run a series of simulations
of a collapsing 1000 $\msol$ clump, changing the
initial conditions namely the initial density and Mach numbers as well as the 
magnetic field \citep{lee2019} by orders of magnitude and found 
 that 
the peak of the stellar mass spectrum is insensitive to these variations in a large range of parameters. 
Moreover they found it to be 
about 5-10 times  the mass, $M_L$, of the 
first hydrostatic Larson core \citep[FHSC][]{larson69,Masunaga98,vaytet2013,Vaytet17}, 
which is the hydrostatic core that forms 
when the dust becomes opaque to its own radiation. To establish this, the effective equation of state, which 
describes the transition from the isothermal to the adiabatic phase, has been varied and it has been found that 
the peak of the IMF occurs at a mass that is proportional to $M_L$. 
The {\it advantage} of this scheme is that it naturally explains the weak  variations of the IMF, at least for 
its peak. The latter is a mere consequence of the physics of the FHSC and its surrounding 
collapsing envelope to be nearly independent of the large scale conditions. 

 In paper II, it has been suggested that the origin
of the factor of 5-10 is due to further accretion from the envelope. Eventually it is limited by the presence of 
nearby fragments which shield, and compete for, further accretion.  
The presence of the nearby fragments is regulated on one hand by the tidal forces induced by the Larson core and
the gas envelope, and  
on the other hand by the density fluctuations arising within its 
surrounding accreting envelope. 
Qualitatively, the mechanism we propose can be summarized as follows. 
Given that  the FHSC is adiabatic, it is clear that fragmentation 
 stops at least up to the point where the second collapse restarts \citep[during which in principle 
tight binaries could possibly form, see e.g.][]{bate1998,machida2008}. Since $M_L$ is the minimum mass to collapse, it seems natural that the peak of the IMF  
should be larger than this value. The infalling gas accumulates and piles up  until a mass of at least  $M_L$
has been accumulated and it sounds unavoidable that further accretion will proceed. Yet it seems 
plausible that the hydrostatic core is surrounded by a mass of several times $M_L$ and that one needs 
to go beyond this mass to find a new hydrostatic core. 

This mechanism, which is largely based on i) a drastic change of the effective equation of state and 
ii) the influence of other objects, put together two ideas that have been proposed earlier.
First  the role played by thermodynamics \citep{jappsen2005} although it emphasizes 
that the important change of the 
effective adiabatic equation of state exponent  is between below and above $4/3$ \citep[][consider the change between 0.7 and 1.1]{jappsen2005}. Second the role played by the fragmentation induced starvation envisioned by 
\citet{peters2011} in which the fragmentation around massive stars limits their growth.   
It presents also  similarities with the competitive accretion scenario advocated by 
\citep{bonnell2001,bonnell2006} though originally this scenario has been proposed to explain the slope of the IMF and the formation 
of the most massive stars rather than the peak of the stellar distribution.

In the present paper we propose a model, which attempts to quantify this picture and
 to estimate  the  mass at which the peak of the  mass spectrum occurs. 
The model consists in counting the density fluctuations in the vicinity of an 
existing protostar that are sufficiently high to be gravitationally unstable
and have a mass that is equal to at least $M_L$. From the probability of finding 
a certain number of fragments, we can infer the typical mass that is unshielded 
and therefore available for accretion into the central object.  
To constrain and test the model, we have performed new simulations $i)$ to
check that the peak is still located at about $5-10 \times M_L$, where 
$M_L$ is the mass of the FHSC, when different initial conditions are explored  $ii)$ to measure the statistics of the 
flow in the vicinity of existing sink particles, $iii)$ to perform a direct test 
of the importance of the fragmentation at few hundreds of AU from an existing protostars in 
setting the peak of the stellar mass function.

The plan of the paper is as follows.
In the second section we present the  numerical simulations, which uses two different setups.
We perform several statistical estimate including numbers of sink neighbours, Mach numbers
and density distributions, cores properties.   
In the third section, we detail the assumptions of the model and the mathematical formalism.
We also compare the predictions of the model with the simulation results. 
The fourth section concludes the paper. 
It is complemented by a long appendix in which the physical properties of the 
FHSC are discussed and simple orders of magnitude are obtained as well as
a discussion on the surface term of the virial theorem.


\setlength{\unitlength}{1cm}
\begin{figure*}
\begin{picture} (0,14)
\put(0,7){\includegraphics[width=9.5cm]{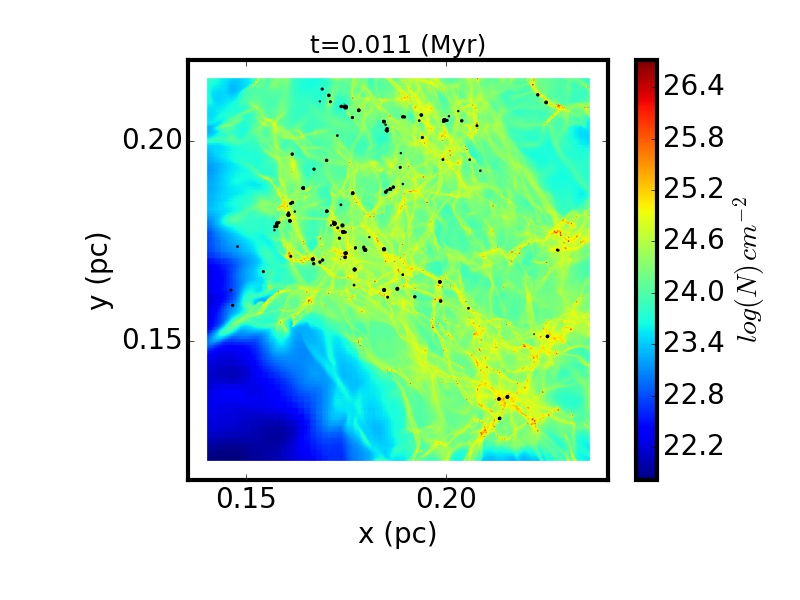}}  
\put(0,0){\includegraphics[width=9.5cm]{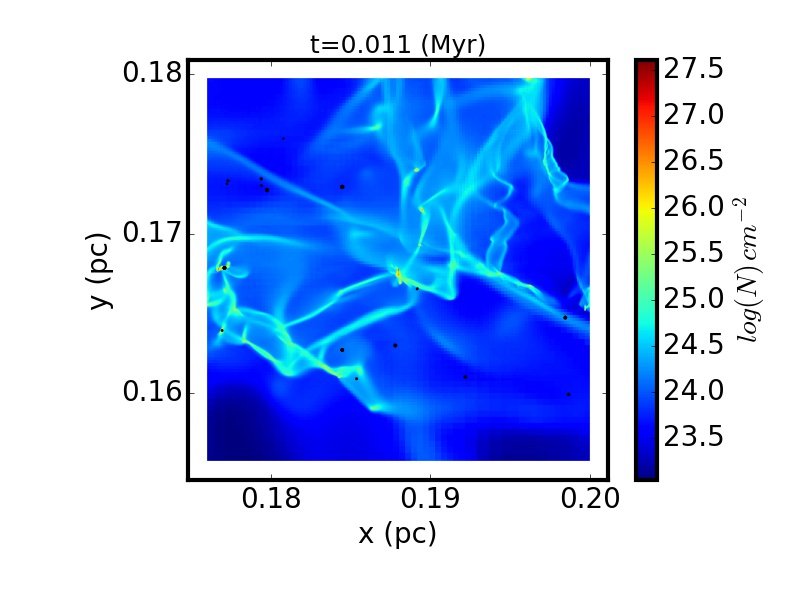}}  
\put(9,7){\includegraphics[width=9.5cm]{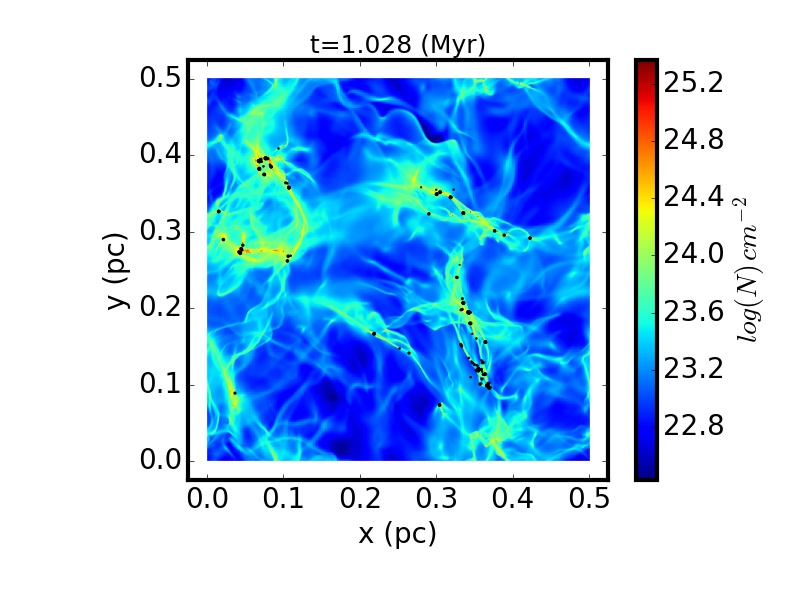}}  
\put(9,0){\includegraphics[width=9.5cm]{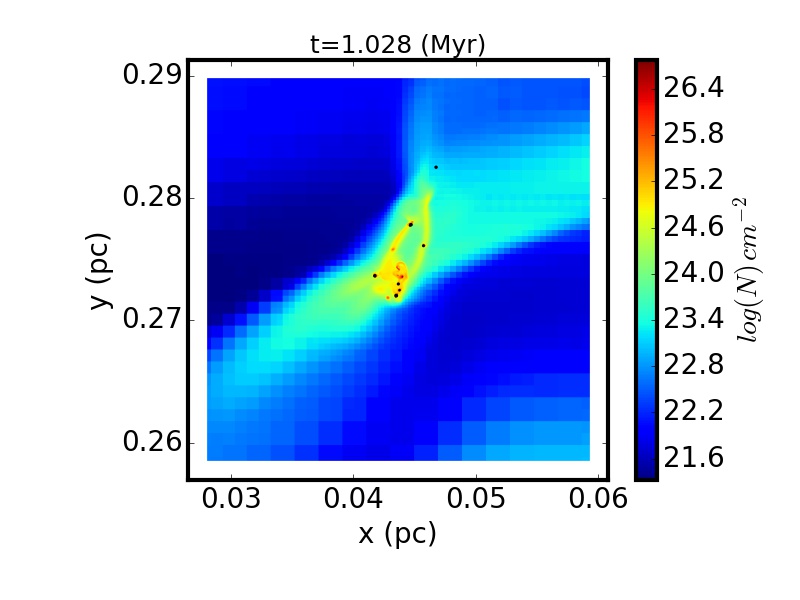}}  
\put(2,13.6){run 1}
\put(11,13.6){run 2}
\end{picture}
\caption{Column density of run1 (left) and run2 (right). Top panels show  most of the whole cloud while 
bottom panels represent zoomed regions.  The black dots represent the sink 
particles aiming to model the stars.}
\label{coldens}
\end{figure*}

\section{Numerical simulations}

\subsection{Numerical methods and setup}

The simulations were run with the adaptive mesh refinement (AMR) magnetohydrodynamics (MHD) code RAMSES \citep{Teyssier02,Fromang06}
and using on a base grid of $(256)^3$  cells, 6 to 7 AMR levels have been added leading in the 
first case to 
an effective resolution of 16384 and a spatial 
resolution of about 4 AU. The Jeans length is resolved with at least 10 points.

\subsubsection{Initial conditions}
In order to verify that our results are not too dependent of a particular choice of initial conditions, 
we have performed  two types of runs. 

 Run 1 considers a spherical cloud with uniform density initially, 
in which turbulence has been added (the fluctuations have random phases and a velocity powerspectrum
which follows the classical powerlaw with a -11/3 exponent) and is freely decaying.
the initial conditions consist in a cloud of $10^3 M_\odot$ which has initially a radius of about 0.1 pc corresponding to an 
initial density of 5$\times 10^6$ cm$^{-3}$. The initial Mach number is about 10. 
The number of AMR levels being used is 14 leading to an 
effective resolution of about 4 AU. As explained below, this run is complemented by two more runs which are 
identical except that the formation of new sink particles is prevented 140 (run1a) and 280 AU (run1b)
around already existing sink particles. 

Run 2 consists in a periodic box in which turbulent forcing is applied for 
wave numbers $k=1-3$. The size of the box is 
0.5 pc, the total mass is also equal to $10^3 M_\odot$ and this corresponds to a density 
of about $10^5$ cm$^{-3}$. The turbulence is driven during 1 Myr before gravity is applied.
By this time the velocity dispersion is about 2 km s$^{-1}$ which corresponds to a 
Mach number of about 10.
The number of AMR levels being used is 15 leading to an effective resolution of about 4 AU.
Note that we have also performed a run with a density 8 times smaller (and a computing box 2 times larger, i.e.
of 1 pc) and inferred very similar results. Therefore this run is not presented here. 

\subsubsection{Equation of state and sink particles}
As found in paper II, the effective equation of state (eos) is playing a fundamental role in 
setting the peak of the IMF. We adopt the following expression:
\begin{eqnarray}
T = T_0 \left\{1 \!+\! { (n / n _{\ad})^{(\gamma_1-1)} \over 
1+( n / n _{\ad,2})^{(\gamma_1-\gamma_2)}  }  \right\}, 
\label{eq_full_eos}
\end{eqnarray}
where $T_0=10$ K, $n _{\ad}=10 ^{10}$ cm$^{-3}$, $n _{\ad,2}= 3 \times 10 ^{11}$ cm$^{-3}$, 
$\gamma_1=5/3$ and $\gamma _2 = 7/5$. This eos mimics the thermal behaviour of the gas when it becomes
non-adiabatic. 
It is well known that an eos with an adiabatic exponent larger than 4/3 leads to an hydrostatic first core
(see section~\ref{larson_core} for orders of magnitude). As discussed in paper II
with this eos, the mass of the FHSC  is about 0.02-0.03 $M_\odot$. Let us stress that its radius is about 
15 AU, which is larger than the 5 AU usually assumed. This makes the FHSC easier to resolve.

We used the sink particle algorithm from \citet[][see paper II for a brief description]{Bleuler14}.
Sink particles are formed at the highest refinement level at the peak of clumps whose density 
is larger than $n = 10^{11}$ cm$^{-3}$.
The sinks are introduced at a density $n \ge 10^{12}$ cm$^{-3}$.  In addition, the clumps 
must be thermally and virially unstable, they should also be contracting. The accretion onto the sinks
occurs within a sphere of radius $4 dx$, i.e. 4 times the smallest resolution element in the 
simulation. At each timesteps $10 \%$ of the gas within this sphere and above 
the density at which sinks are introduced, is retrieved from the computational cells and 
assigned to the sink. In paper II the value at which the sinks are formed and  accrete  as well as the fraction
of the dense gas that is transfered to the sinks have been 
varied. 
It has been found that they do not have a strong impact on the resulting mass spectrum. The influence 
of numerical resolution, to which the sink radius is proportional, has also been checked and 
it has been found that it has no significant influence on the peak position.

\setlength{\unitlength}{1cm}
\begin{figure*}
\begin{picture} (0,7.5)
\put(0,0){\includegraphics[width=8.7cm]{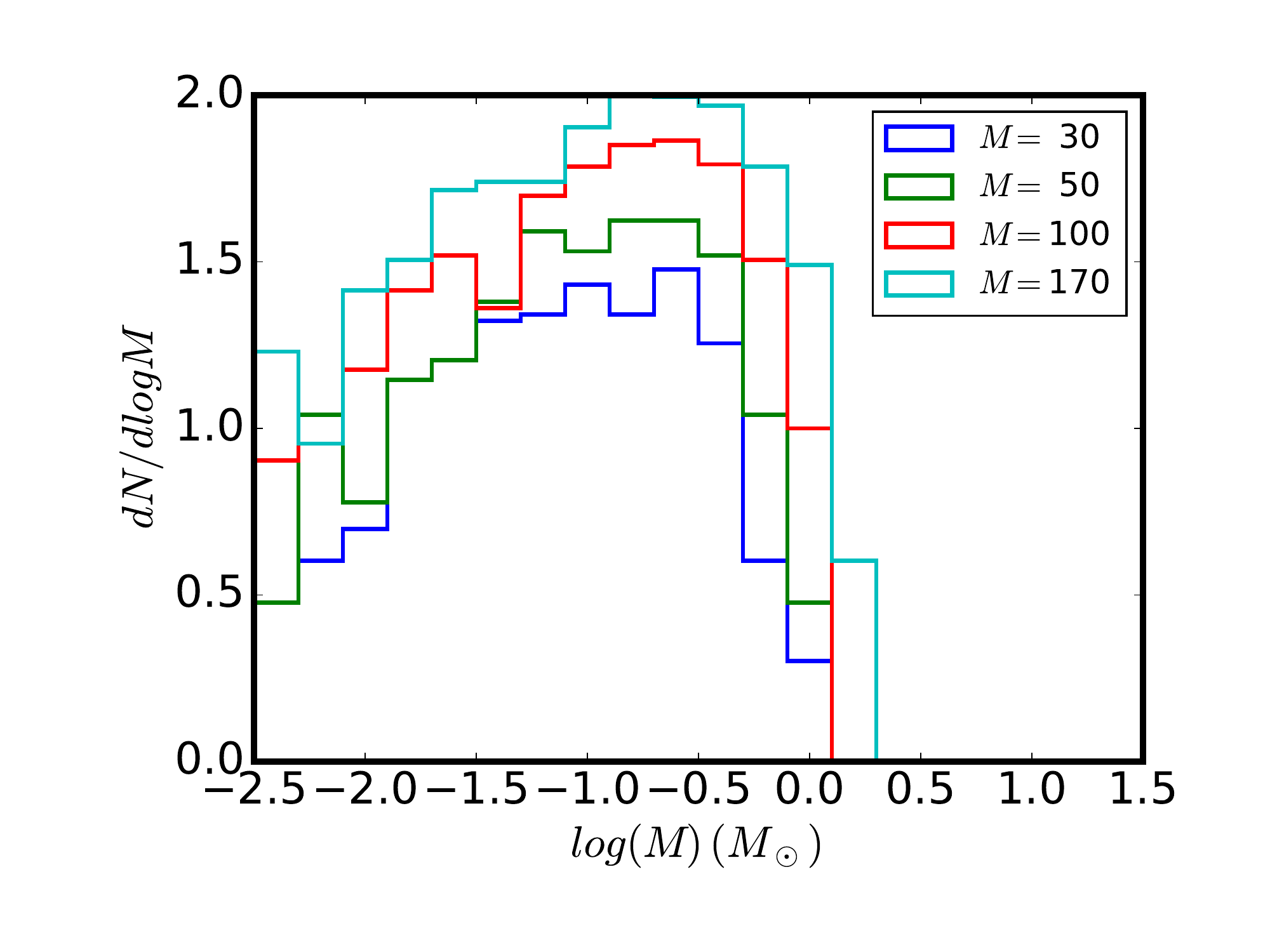}}  
\put(8,0){\includegraphics[width=8.7cm]{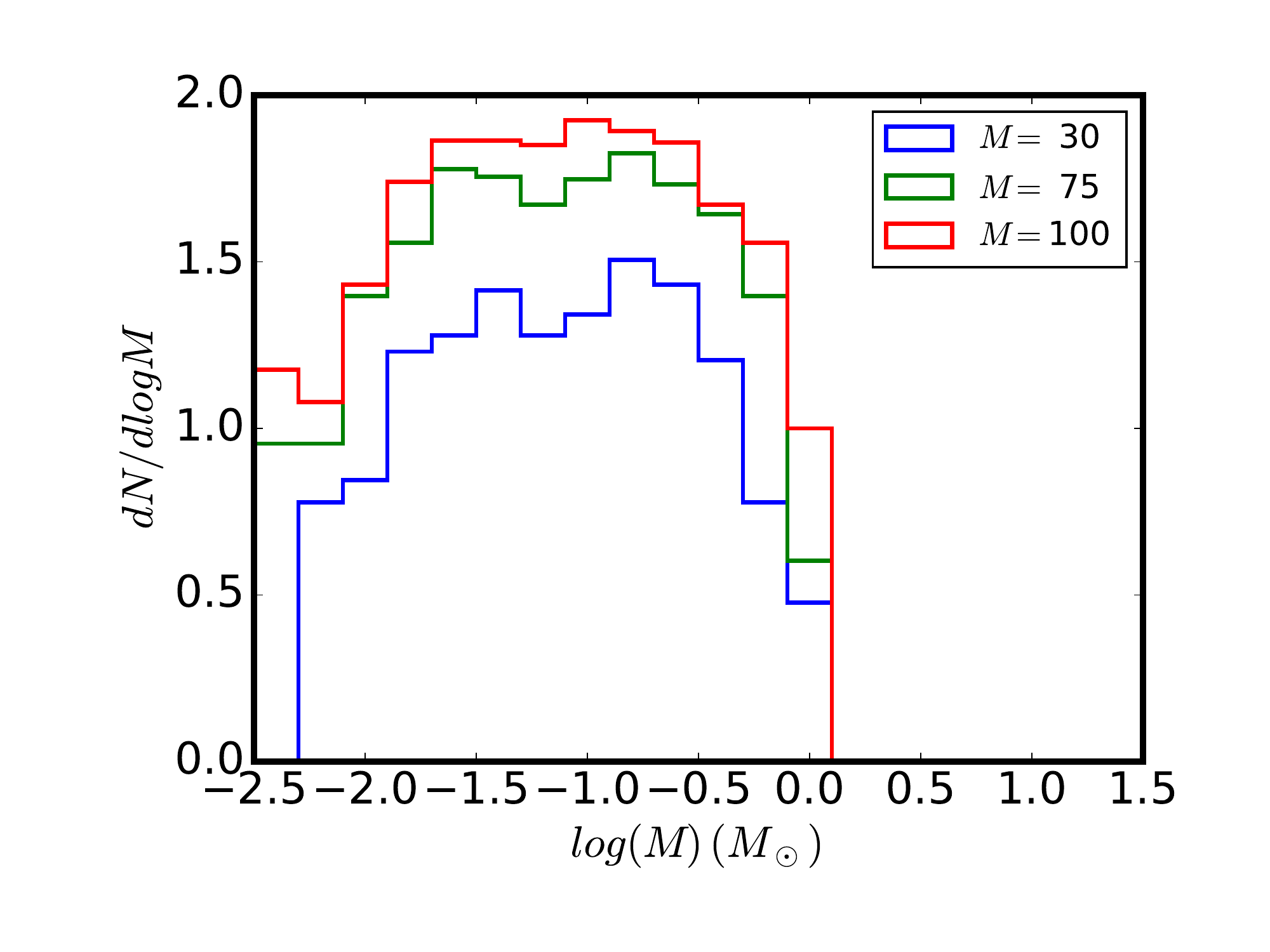}}  
\put(2,6.){run 1}
\put(10,6){run 2}
\end{picture}
\caption{Mass spectrum at various times for run 1 (left) and run2 (right).
The total mass accreted by the sink particles (expressed in solar mass) is indicated in the legend.
 The distribution peaks at about 0.2 $M _\odot$ for run1 and 0.1 $M _\odot$ for run2.}
\label{run_IMF}
\end{figure*}

\setlength{\unitlength}{1cm}
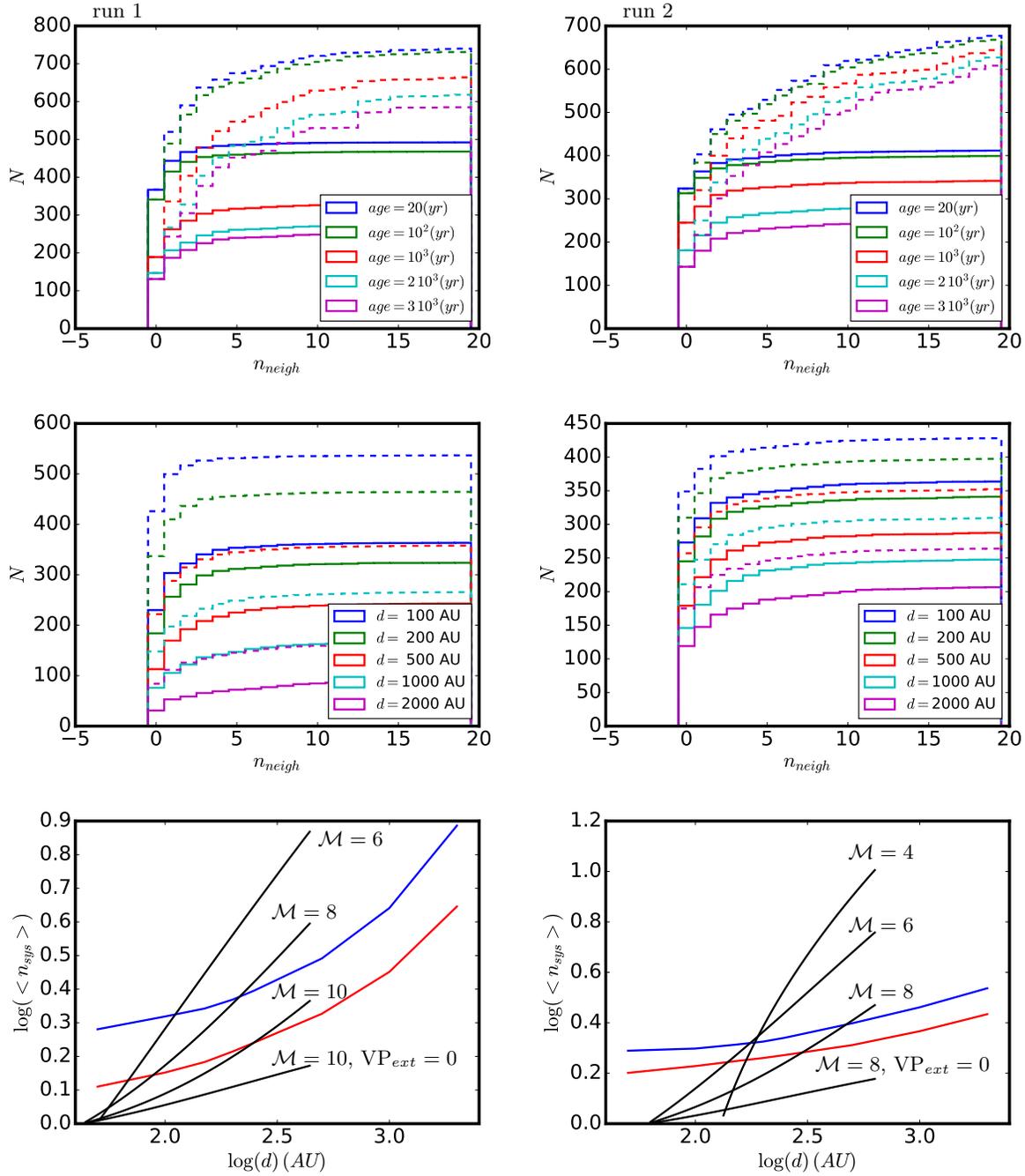
\begin{figure*}
\begin{picture} (0,19)
\put(0.,12){\includegraphics[width=8.7cm]{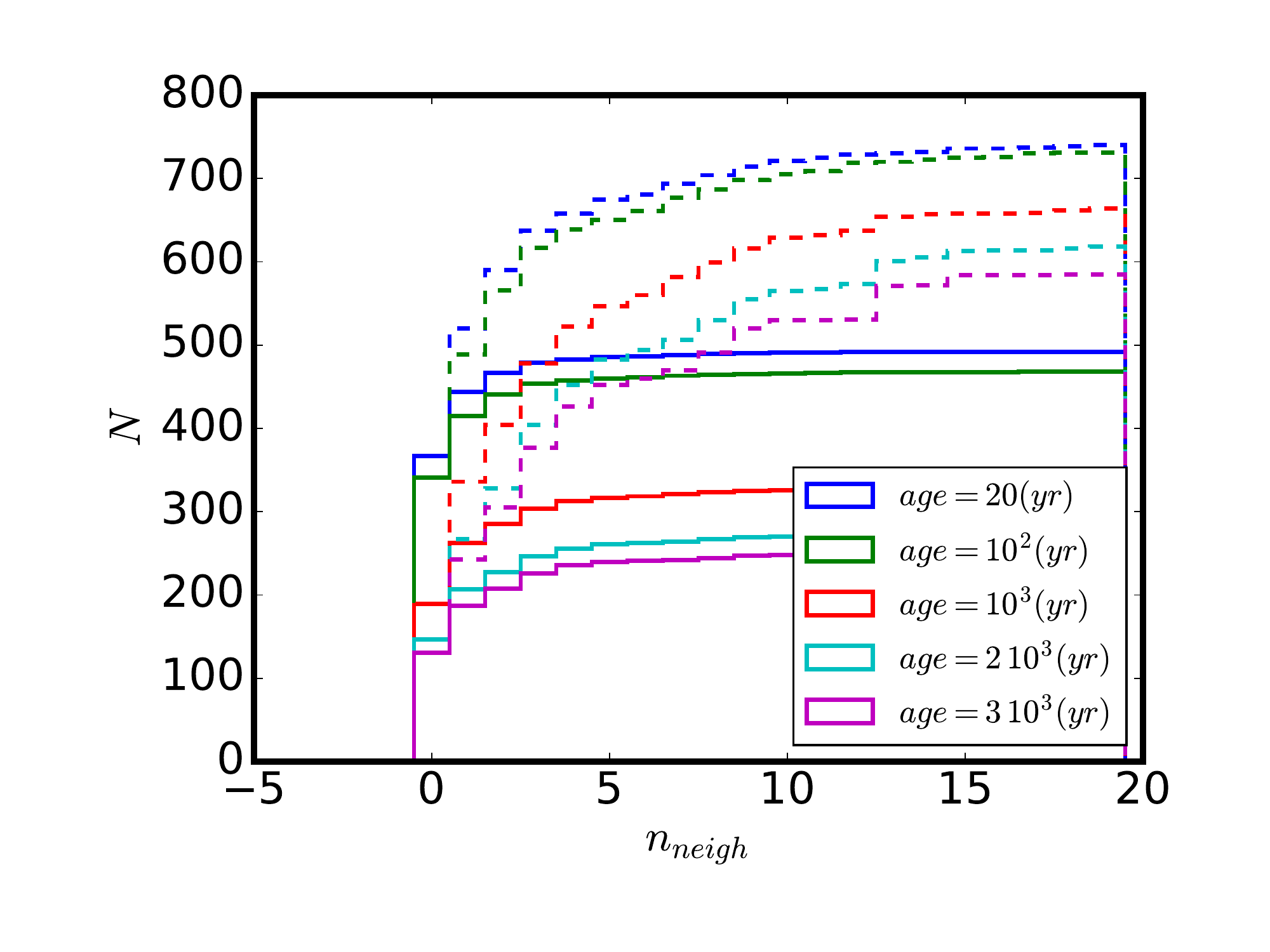}}  
\put(0.,6){\includegraphics[width=8.7cm]{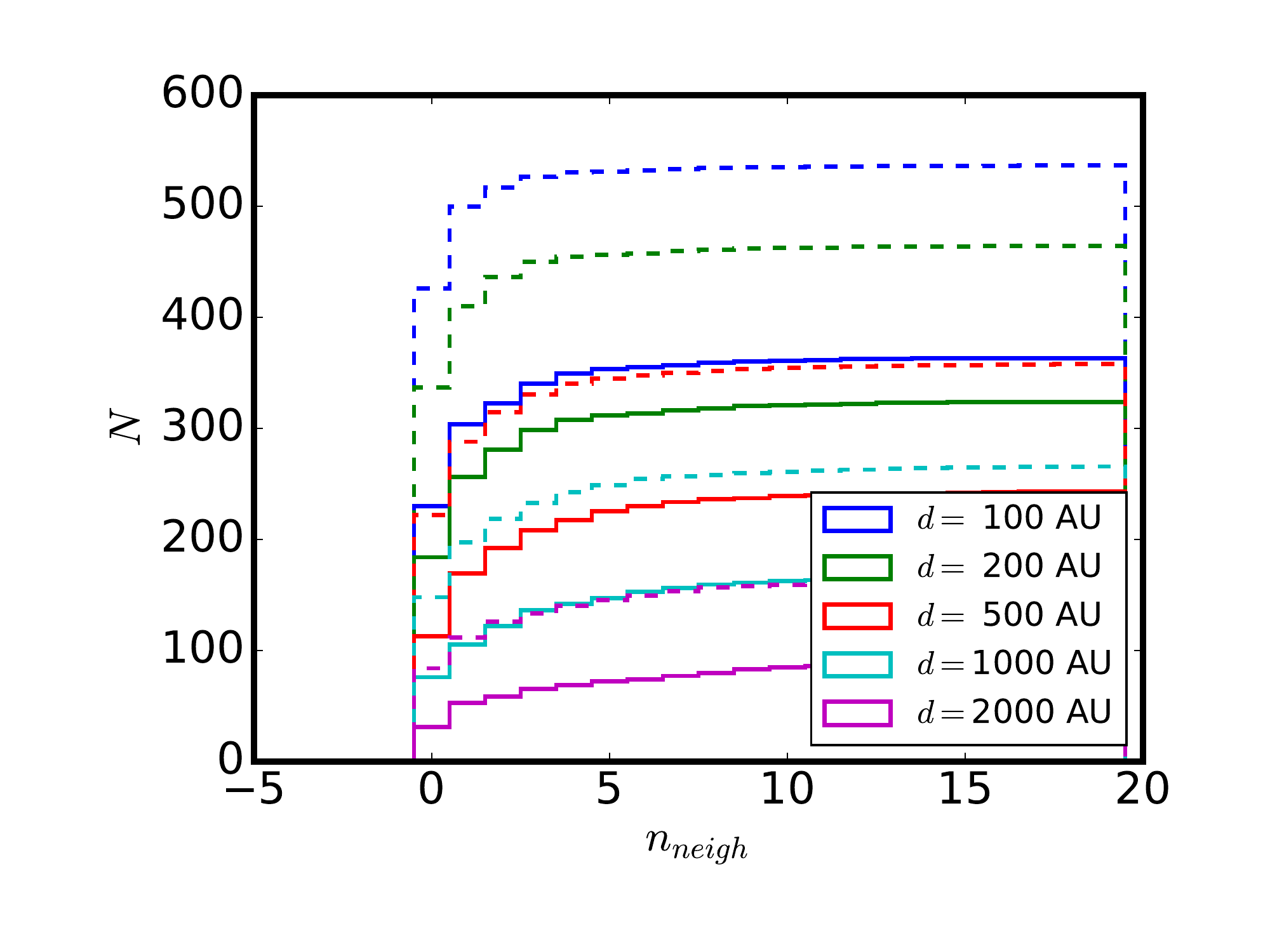}}  
\put(0.,0){\includegraphics[width=8.7cm]{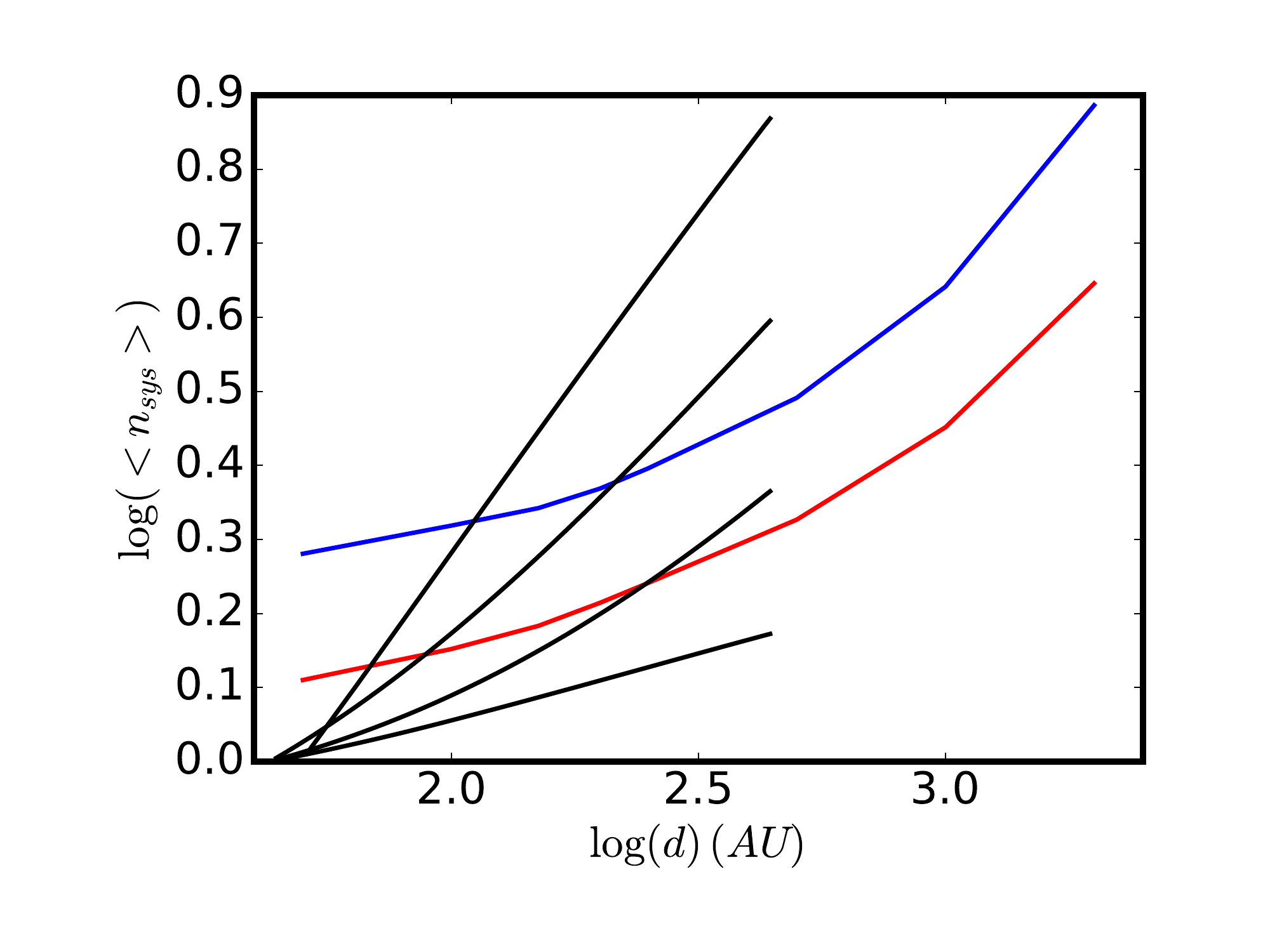}}  
\put(8,12){\includegraphics[width=8.7cm]{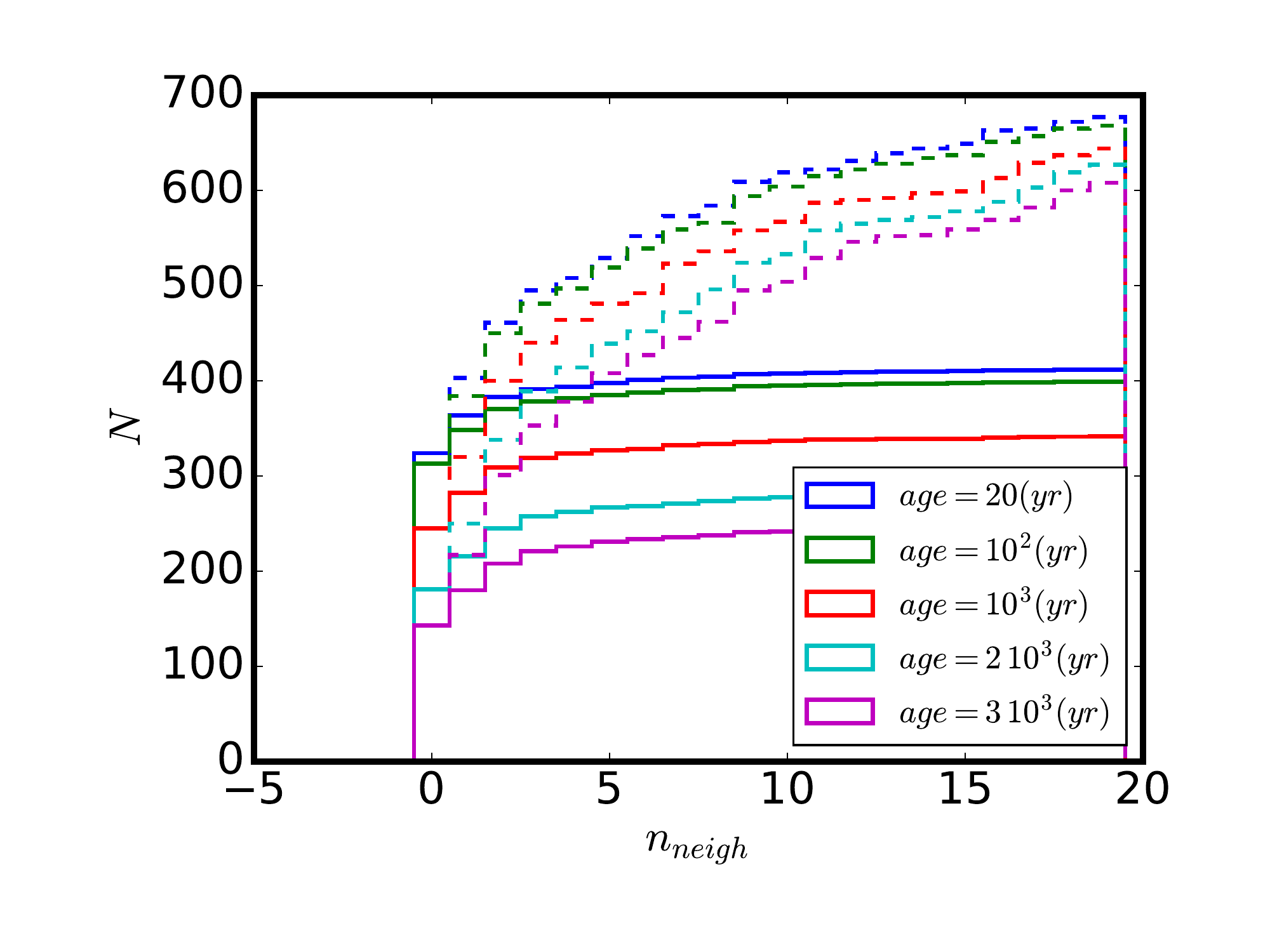}}  
\put(8,6){\includegraphics[width=8.7cm]{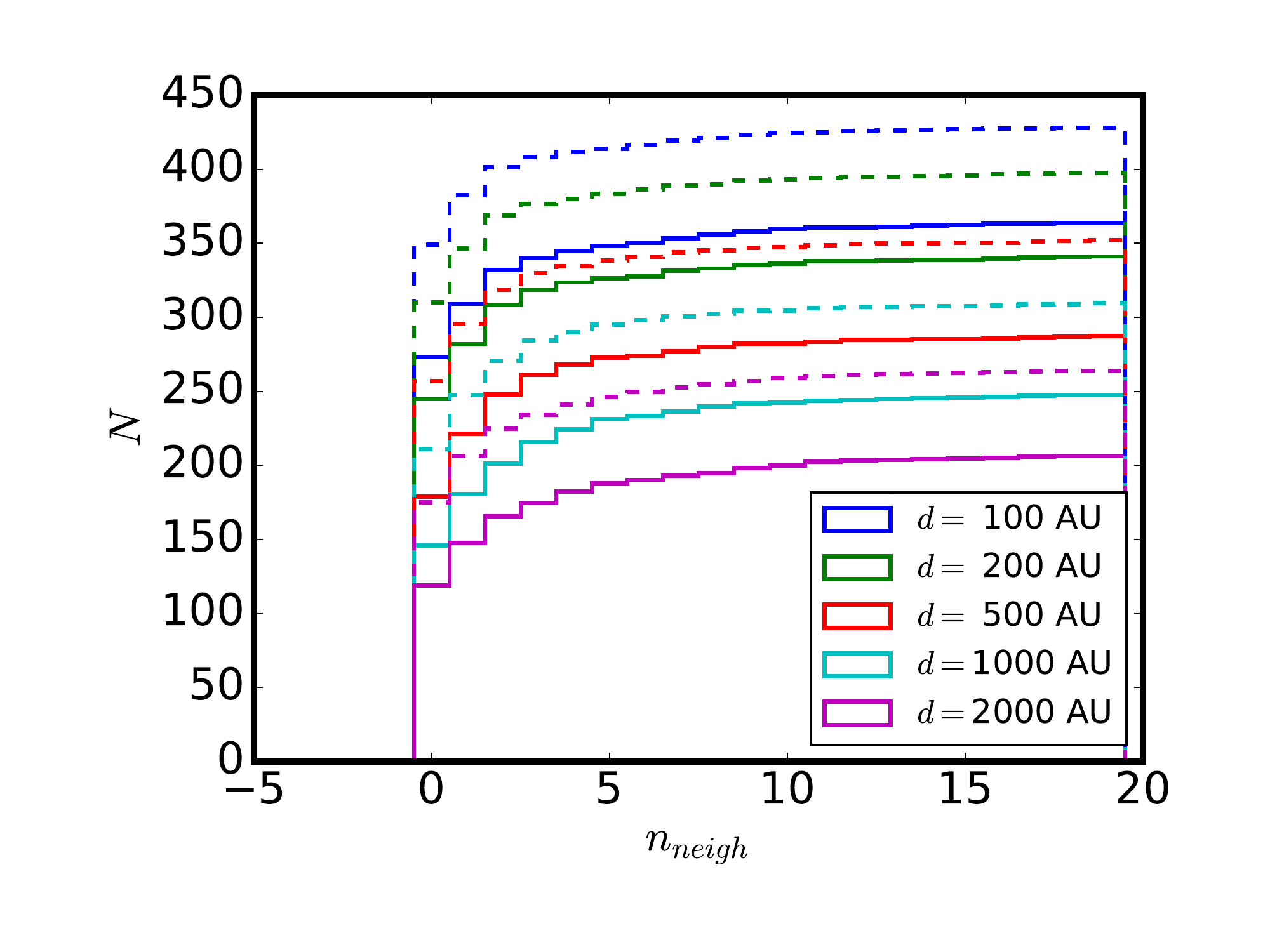}}  
\put(8,0){\includegraphics[width=8.7cm]{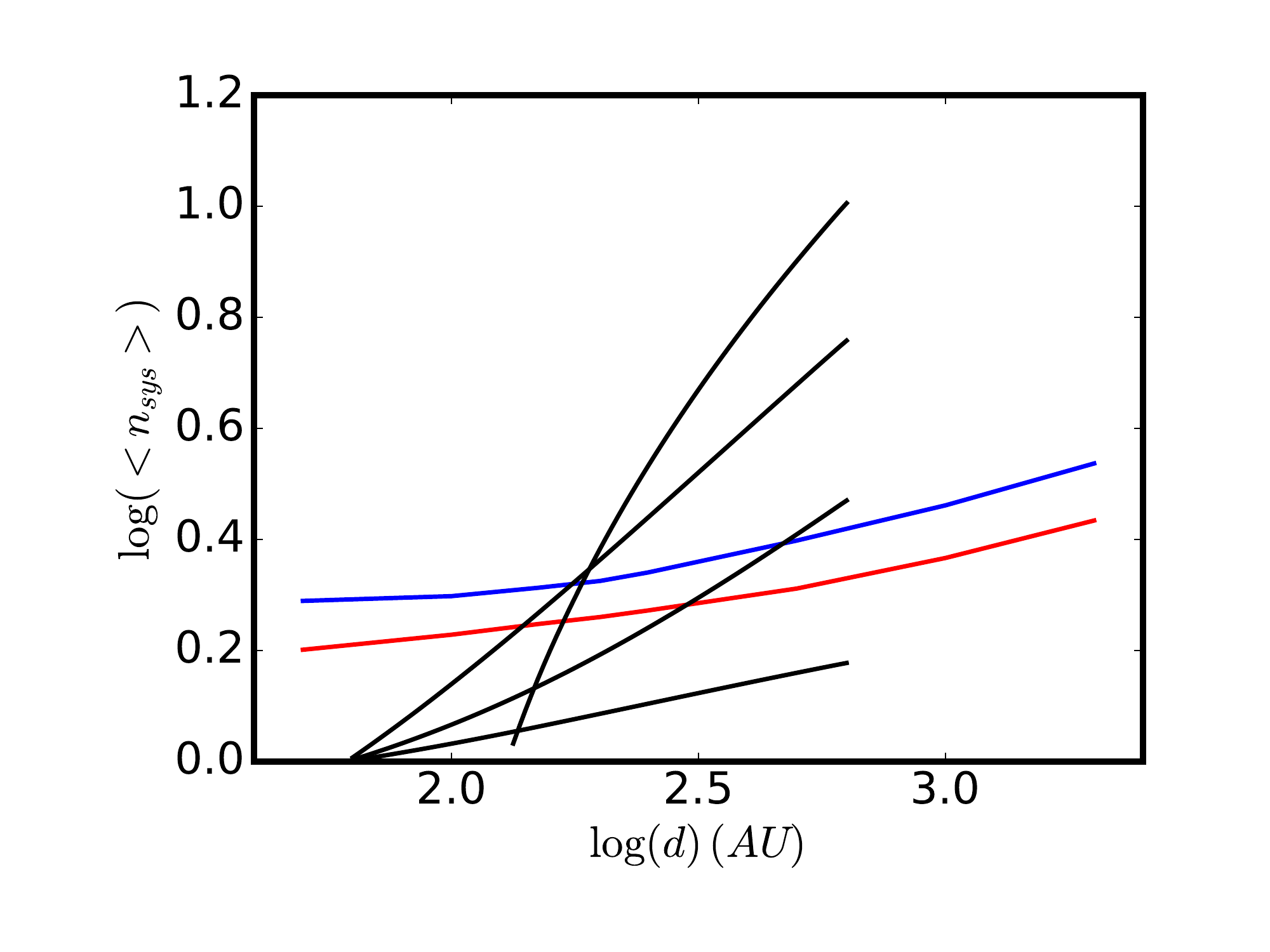}}  
\put(2,18){run 1}
\put(10,18){run 2}
\put(4.7,2.2){${\mathcal M}=10$, ${\rm VP}_{ext}=0$}
\put(4.7,3.2){${\mathcal M}=10$}
\put(4.7,4.4){${\mathcal M}=8$}
\put(5.4,5.5){${\mathcal M}=6$}
\put(12.9,2.1){${\mathcal M}=8$, ${\rm VP}_{ext}=0$}
\put(13.4,3.2){${\mathcal M}=8$}
\put(13.4,4.2){${\mathcal M}=6$}
\put(13.4,5.3){${\mathcal M}=4$}
\end{picture}
\caption{Neighbour statistics. Left is for run 1 and right for run 2. 
Top panels give the cumulative histogram
of the number of {\it systems}, $N_{sys}$, (solid lines) of size 200 AU which contain  $n_{sys}=n_{neigh}+1$ objects. 
The dashed lines show the number of
sinks ($N_{neigh}=N_{sys} n_{sys}$) having $n_{neigh}$ neighbours 
located at a distance of less than 200 AU
at various times  after their formation.  
Middle panels provide 
the cumulative histogram of the number of systems possessing $n_{neigh}+1$ objects 
at various distance and at time t=10$^2$ yr (dotted line) and at time t=10$^3$ yr (solid line).
 Bottom panels display the 
mean system number, $n_{sys}$, as a function of distance. Red curves are times 
shorter than 10$^2$ yr while blue ones are for $t<10^3$ yr.  The black lines 
show the result of the analytical model inferred in Sect.~\ref{sec_analyt} for various 
Mach numbers (see text in section~\ref{comparison}). Good agreement between the model and the 
simulations for the highest Mach numbers measured is obtained. When the external pressure terms
are not taken into account the number of objects is below the numbers inferred in the simulations.}
\label{run1_neigh}
\end{figure*}

\subsection{Qualitative description}
To illustrate the simulation results, the column density is portrayed 
in Fig.~\ref{coldens}. Left is run1 and right run2. The sink particles have 
been overlaid and represented by the dark dots. In both runs 
the dense gas is organised in dense filaments. The distributions of the filaments 
is however different. In run1, they tend to be more numerous and more uniformly 
distributed than in run2, which is organised in few massive objects in which most 
of the star formation is taking place. 
 The simulations have been run respectivelly during a time of 0.013 Myr and 1.04 Myr 
(which corresponds to 0.04 Myr after gravity is switched on). In the two cases, 
this is about 1 freefall time of the dense gas.

\subsection{Sink mass  distributions}
Figure~\ref{run_IMF} displays the mass distribution of the sink particles for run1 (left) and 
run2 (right) at several timesteps which correspond to various total accreted mass. 
In run1  a marked peak at $\simeq$0.2 M$_\odot$ is inferred. At higher mass, the distribution is 
compatible with a powerlaw $M^{-0.75}$ as found in paper I. In run2, the peak is less clearly
pronounced because there are more low mass objects than in run1, the corresponding  distribution 
tends to be flatter. There is nevertheless also a peak which located at 0.1-0.2 M$_\odot$. 
At higher mass, the distribution again is broadly compatible with a powerlaw $M^{-0.75}$. 

This is in good agreement with the results obtained  in paper I and II. 

 At the end of the calculations,
 the fraction of gas that has been accreted into the stars is about 17$\%$ for run1 and 
$10\%$ for run2. The total numbers of sinks formed are respectivelly 762 and 723. 
While these numbers are sufficient to produce reasonable statistics, the question as to whether 
the mass spectrum may vary at later time arises. However Fig.~\ref{run_IMF} reveals that the shape of 
the mass spectrum, and particularly, the peak, does not vary very significantly between the 
times where 30 $M_\odot$ and 170 $M_\odot$ has been accreted. In paper I, we were able to go up to 
the point where 300 $M_\odot$ of gas have been accreted. Therefore, we expect this integration time 
to be sufficient to  estimate the mass spectrum. Another related issue, is the stellar feedback, that should 
become important once enough mass has been converted into stars, and that may even disperse the clouds 
before all the gas is turned into stars. 

\subsection{Sink neighbour distributions}
We now turn to the analysis of the sink neighbor distribution. Our goal here is to investigate 
to what extent the presence of sinks in the immediate vicinity may influence the peak 
of the stellar mass function as proposed in paper II. To address this question, for each sink we have calculated 
the number of neighbours, $n_{neigh}$, located below a series of specific distances and for various sink ages. 
The statistics are obtained by analysing the sink configurations during the entire simulation. 
 In practice, we consider all sinks formed in the simulations and at each timestep we 
calculate the number of neighbours lying within a series of specific distances. To define the number 
of neighbours, we simply select all timesteps for which the corresponding objects have an age below 
a specific value and take the maximum of neighbours found. In practice, we do not find much fluctuations 
in the number of neighbours. Once its maximum value is reached, it rarely decreases.
As it is more common to draw the statistics of the number of systems 
(that is to say the number of groups of neighboring stars)
 and also because this is what the 
analytical model developed below produces,  we mainly present the statistics of the number of systems of
size $n_{sys}=n_{neigh}+1$. 
The number, $N_{sys}$, of systems with $n_{sys}$ objects, is related to the number, $N_{neigh}$ of sinks
with $n_{neigh}$ numbers through $N_{sys}=N_{neigh} / (n_{neigh}+1)$.
Note that the systems we are interested in here may not be necessarily gravitationally bound (although it is the case for 
most of them), because we specifically address the question of the influence of fragmentation onto the 
sink mean mass. 

The results are portrayed in Fig.~\ref{run1_neigh}.
Top panels show the cumulative histogram
of  system numbers ($N_{sys}$, solid lines) and sink numbers ($N_{neigh}$, dashed lines) 
of various ages with (up to) $n_{neigh}=n_{sys}-1$ neighbours  at a distance below 200 AU.
As can be seen at early time (t $\le$ 100 yr) more than half of the sinks have 0 to 2 neighbours below
200 AU and about 3/4 of the systems are constituted by a single object. 
 Interestingly a fast evolution is visible between t=100 yr and t=1000 yr, especially 
for $n_{neigh}<2$. Indeed, we see that the distance between the curve of age 100 and 
1000 yr is larger than what it is between 2000 and 3000 yr for example.
This indicates that fragmentation in the vicinity of young sinks is 
commonly occurring. A significant number of objects are surrounded by a relatively large number 
of neighbours, $n_{neigh} \ge 10$ but the corresponding system number is small. 
This corresponds to a restricted number of regions where the clustering 
is particularly important. 

 Middle panels display
the cumulative histogram of the number of systems  with $n_{neigh}=n_{sys}-1$ objects
for several distances and at time 10$^2$ yr (dotted line) and at time 10$^3$ yr (solid line). 
As expected the number of 
neighbours increases with the distance, however we clearly see that this increase 
is not very fast and clearly  sublinear with distance. 

To quantify this better, bottom panels show  the 
mean system size, $<n_{sys}>$ as a function of distance. Red curves are for 
 $t \le 10^2$ yr while blue ones are for $t \le 10^3$ yr.  The black curves
correspond to the analytical model and will be discussed later. 
By comparing red and blue curves of run1 (left panel), it is very clear that   $<n_{sys}>$ below 
100 AU is $\le$ 2 at $t=100$ yr and $\simeq$2 at $t=1000$ yr.
This is  similar for run2.
At 1000 AU there is more difference between run1 and run2 with about 4-5 objects for the 
former and $\le$ 3 for the latter.

\setlength{\unitlength}{1cm}
\begin{figure}
\begin{picture} (0,7.5)
\put(0,0){\includegraphics[width=8.7cm]{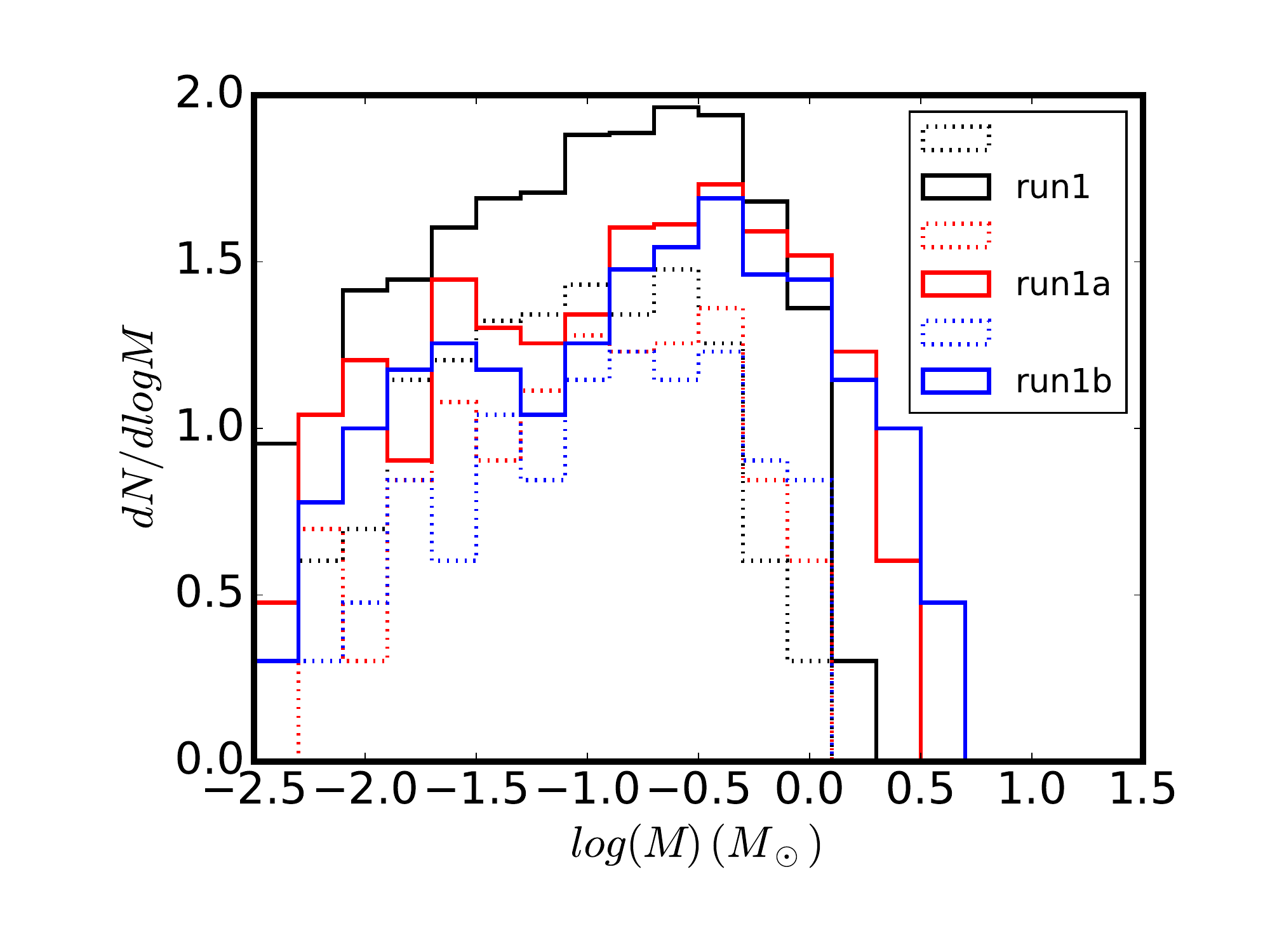}}  
\end{picture}
\caption{Mass spectra for run1, run1a (sinks are forbidden to form below a distance of 140 AU from an existing one) and run1b (sinks are forbidden to form below a distance of  240 AU from an existing sinks). Solid lines are for 
a total sink masses of 150 $M_\odot$ while dashed lines stand for 50 $M_\odot$. 
Clearly the sink distribution shifts  to larger masses from run1 to run1a and run1b. There are altogether 2 times less sinks in run1b than in run1 and their masses are about 2 times larger. This clearly demonstrates that 
fragmentation within the inner hundreds of AU has a drastic impact on the sink/stellar mass distribution. }
\label{run1_IMF_exp}
\end{figure}

\subsection{A simple numerical experiment}
To demonstrate that the fragmentation close to existing objects is critical in setting the peak of the IMF, we seek for 
a clear test. We cannot increase the formation of sinks without changing the EOS, which affects the mass of the 
FHSC and the peak of the stellar mass function as shown in paper II. 
We can however prevent the formation of new sinks below a given distance from 
an existing sink by enforcing it in the simulation. 
We have therefore performed two runs, called respectively run1a and run1b, where sink 
formation is prevented below respectively 140 and 280 AU from any existing sink. 

The resulting mass spectra are portrayed in Fig.~\ref{run1_IMF_exp}. Clearly the sink distributions shift to higher masses 
as the distance from an existing sink, below which new sinks cannot form, increases.  Solid lines are for a total mass of sinks 
equal to 150 $M_\odot$ while dashed ones are for 50 $M_\odot$. Both the peak and the mass of the most 
massive sinks shift by nearly a factor of 2. 
 In run1b there are about two times less 
sinks and the sinks are about two times more massive than in run1. 
These numbers are entirely compatible with the neighbour distributions displayed in Fig.~\ref{run1_neigh} 
which reveal that 1-2 objects on average sit at distance  below 100-200 AU from an existing sink. 

These simulations  demonstrate that 
indeed the close fragmentation that limits the accretion onto most of the objects is determinant for their masses. 
By preventing fragmentation at distance of 100-200 AU from existing sinks, the peak of the stellar mass function shifts 
by a factor of 2 because there is more gas available for accretion. In the presence of more fragments, less gas is 
available and less massive stars form.

\setlength{\unitlength}{1cm}
\begin{figure*}
\begin{picture} (0,19)
\put(0,0){\includegraphics[width=8.7cm]{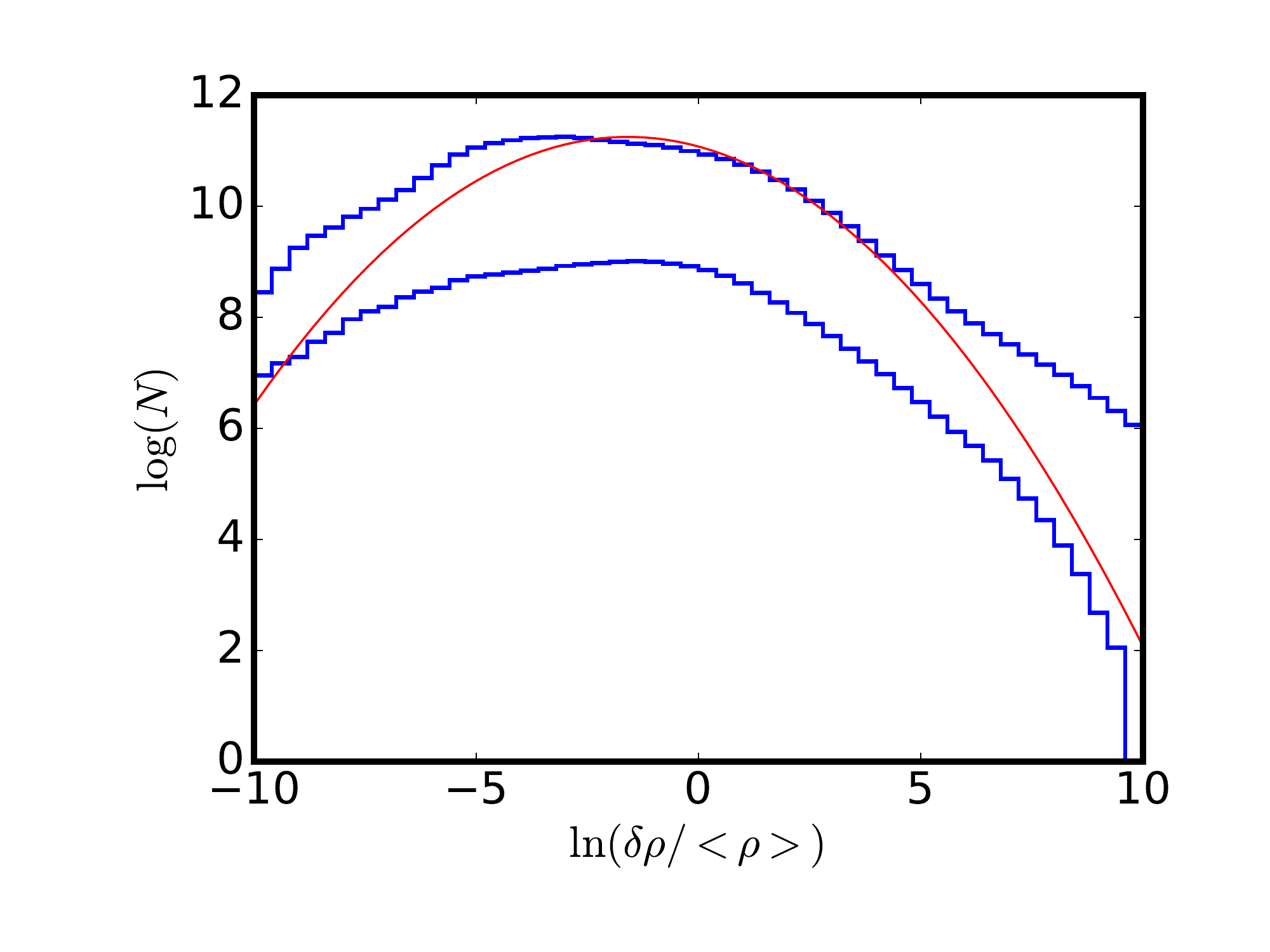}}  
\put(0,12){\includegraphics[width=8.7cm]{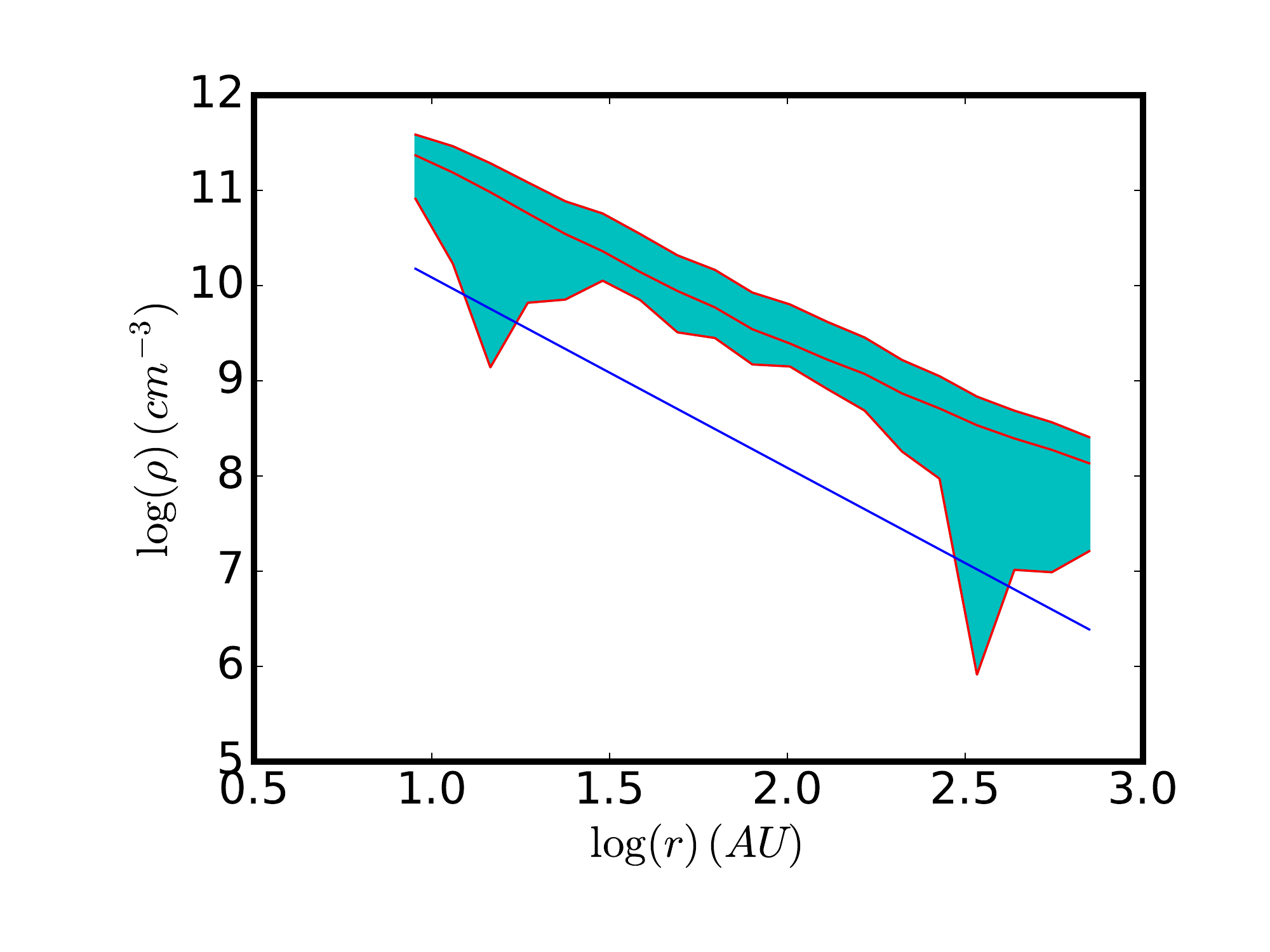}}  
\put(0,6){\includegraphics[width=8.7cm]{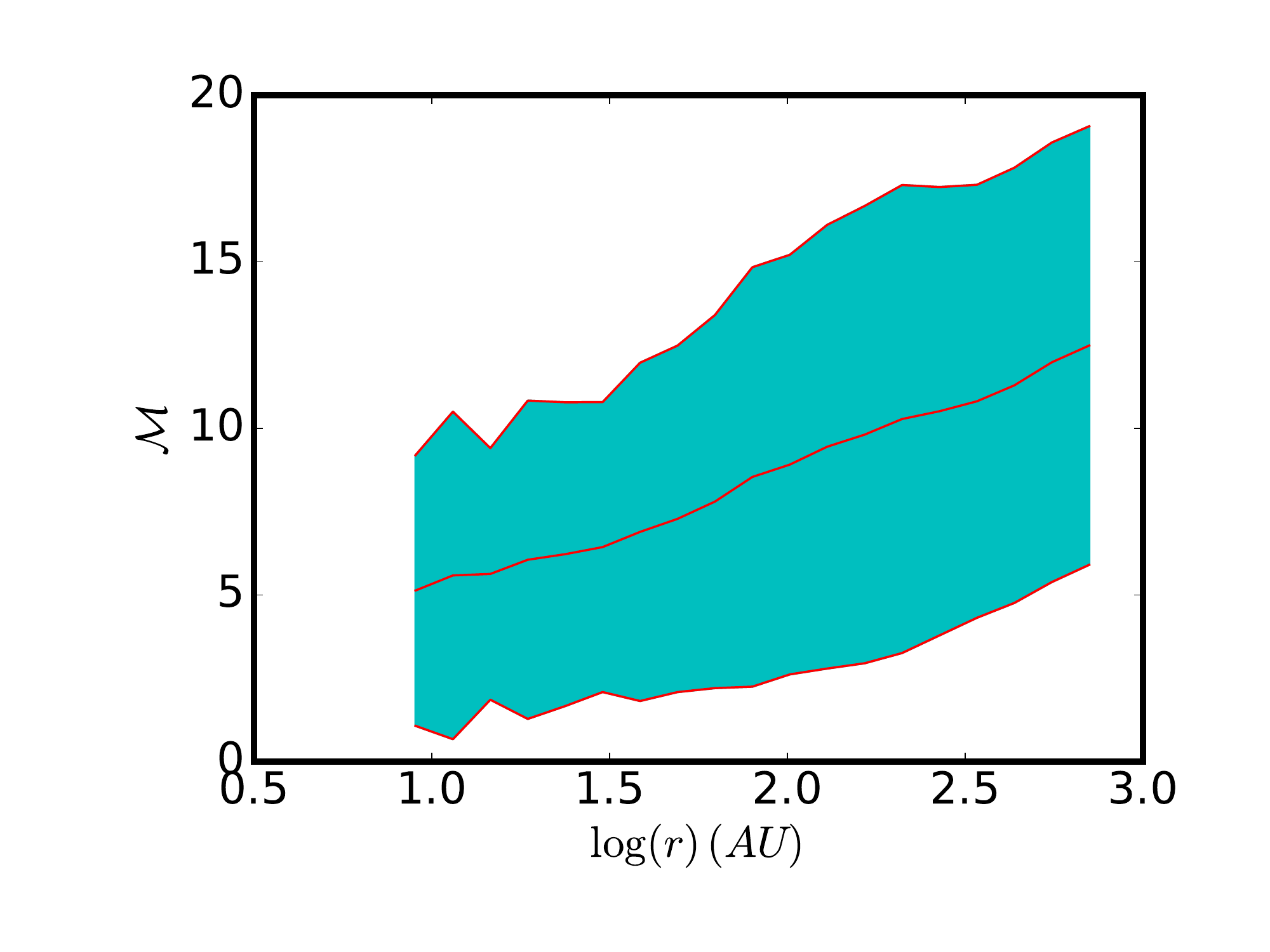}}  
\put(8,0){\includegraphics[width=8.7cm]{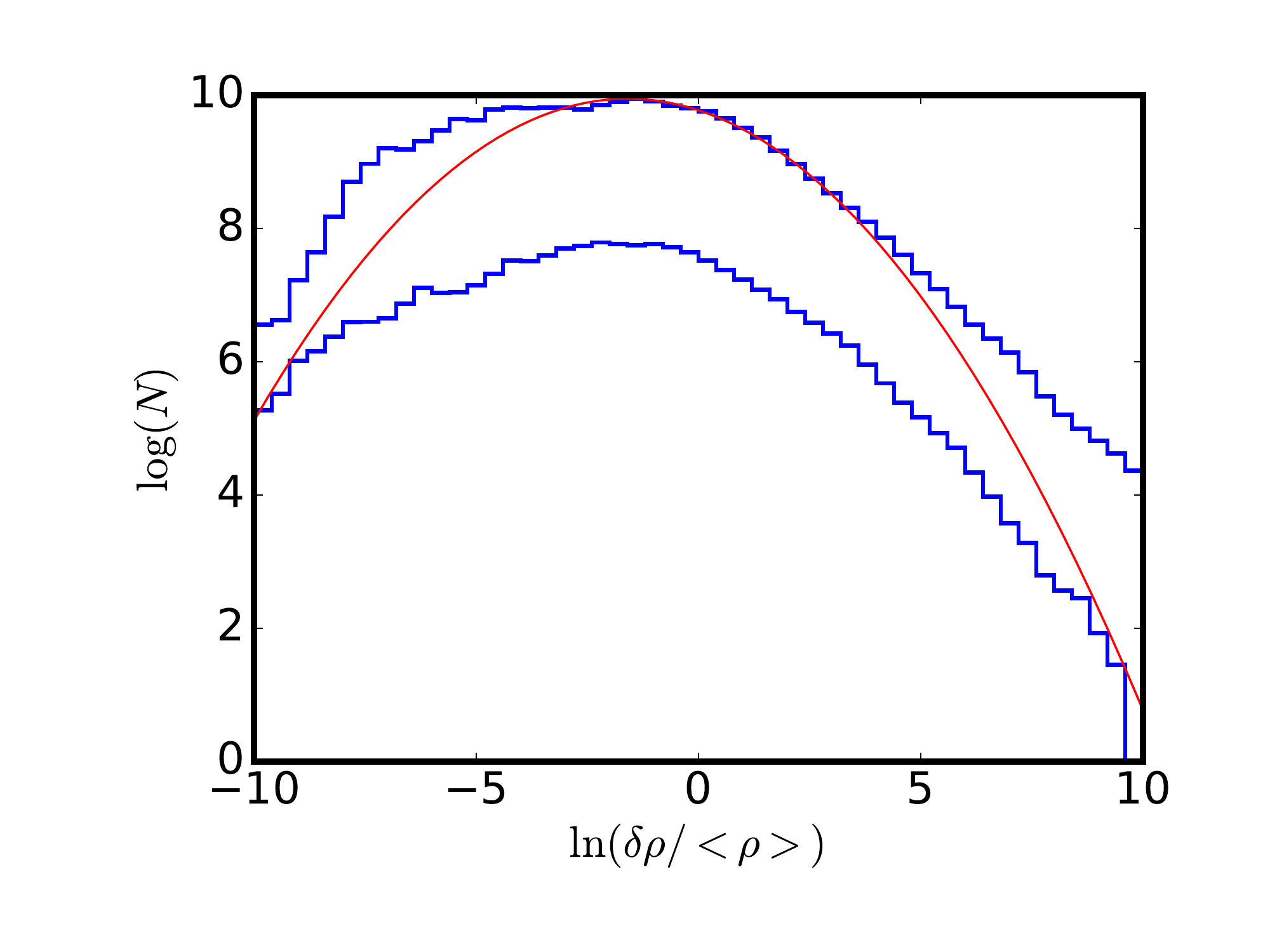}}  
\put(8,12){\includegraphics[width=8.7cm]{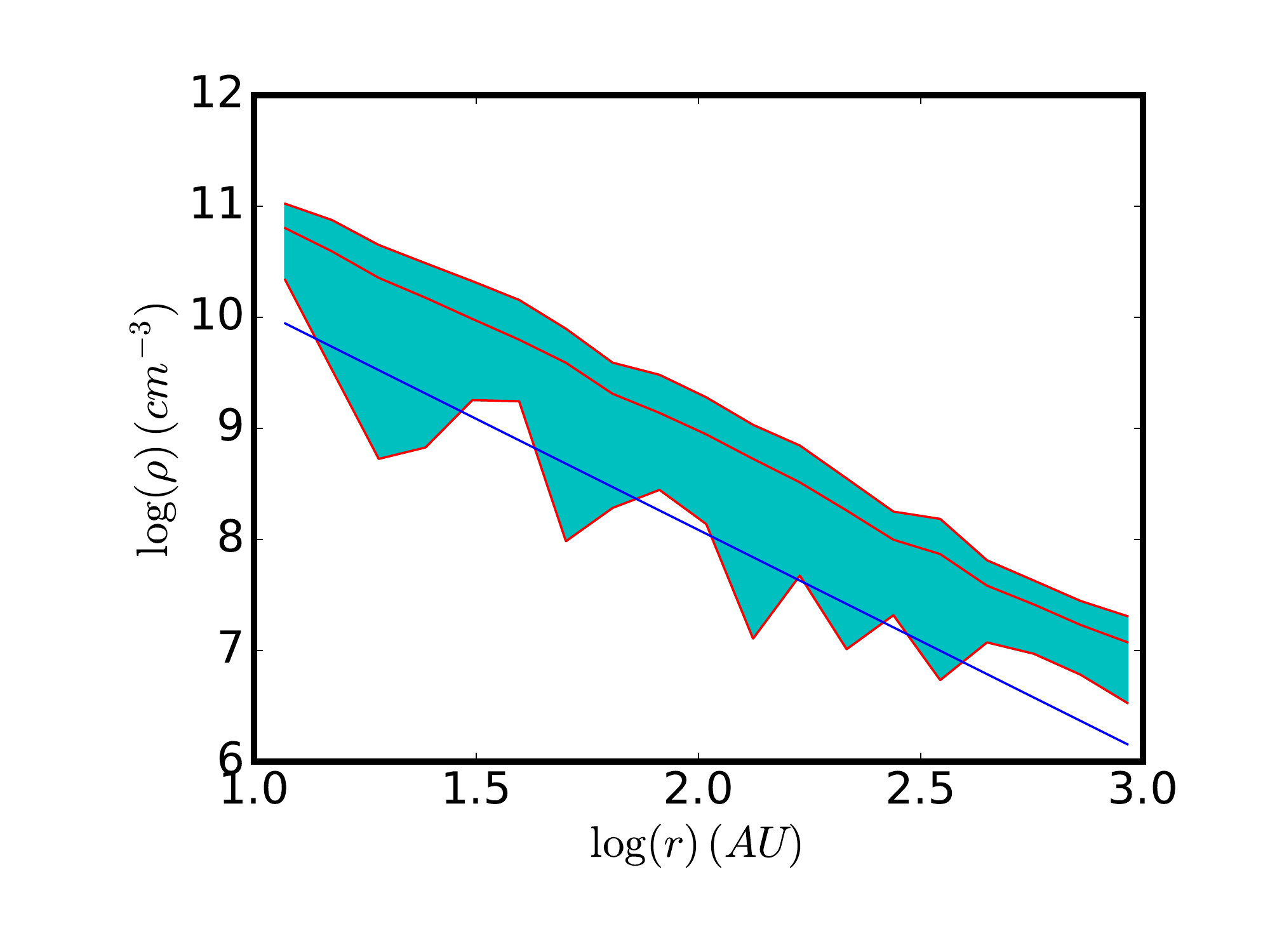}}  
\put(8,6){\includegraphics[width=8.7cm]{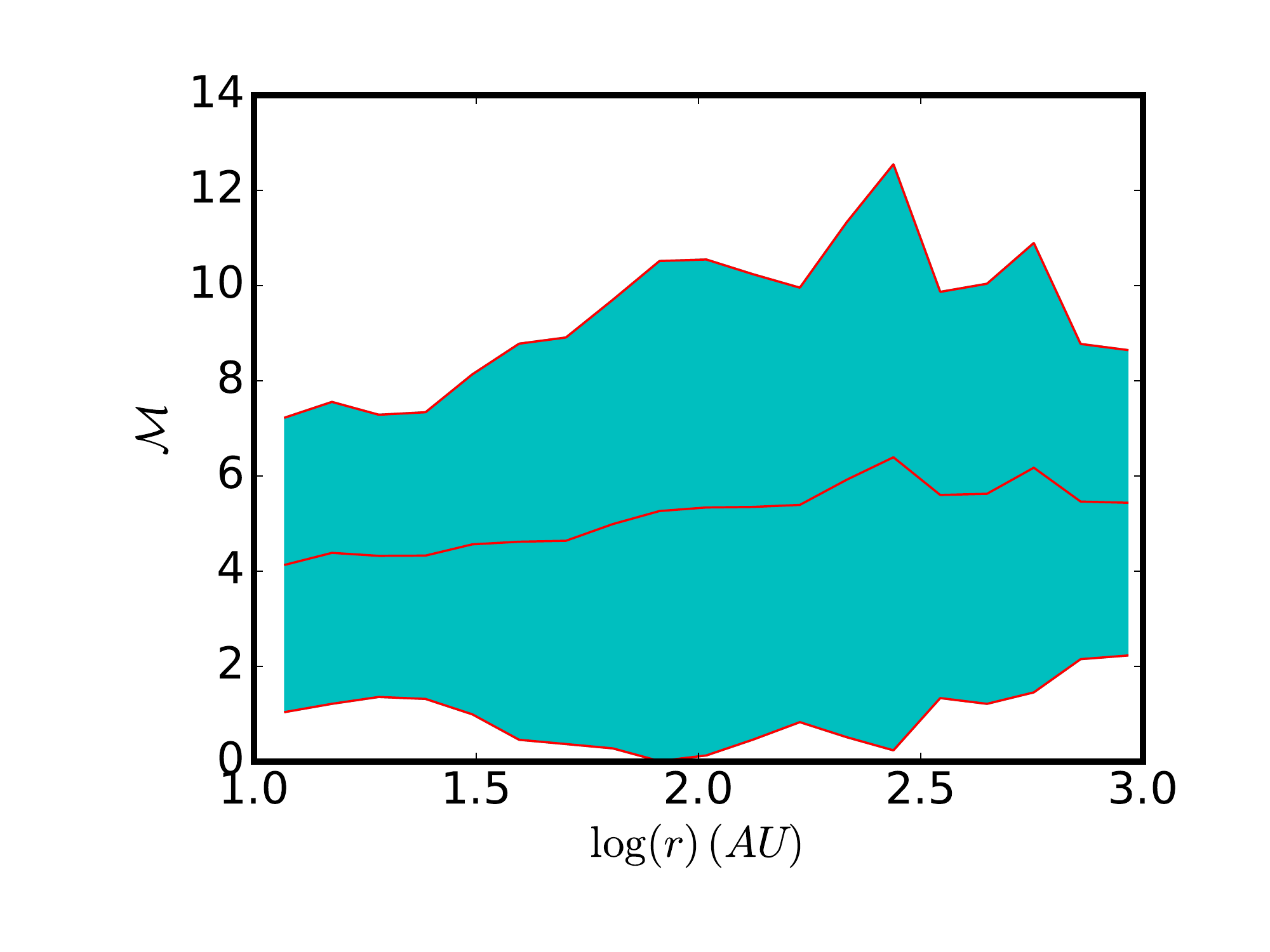}}  
\put(2,18){run 1}
\put(10,18){run 2}
\end{picture}
\caption{Gas statistics around the sinks younger than 1000 yr. Left: run1, right: run2. 
Top panels show the mean density (red line) and 
the mean standard deviation (blue shaded area). 
For reference, the blue line shows the density of the singular isothermal sphere, $n_{\rm SIS} = \rho_{\rm SIS} / m_p$. 
We see that on average the density is about $10 \times \rho_{\rm SIS}$ for run1 and about $7 \times \rho_{\rm SIS}$ for run2. 
Middle panels display the mean Mach numbers, which below 
300 AU is found to be about 5-10 for run1 and 4-5 for run2. 
Bottom panels portray the distribution of the density fluctuations (i.e. with respect to the mean density shown on 
top panel) within a sphere of radius 800 AU (upper curve) and 250 AU (lower curve). 
For reference the red curves represent a log-normal distribution with a width 
$\sigma^2 = \ln (1 + b^2 \mathcal{M}^2)$ with $b=0.8$ and $\mathcal{M}=6$. }
\label{run1_stat}
\end{figure*}

\subsection{Sink environment} 
As fragmentation around existing objects appears to be determinant, we need to characterise the physical conditions that prevail 
around newly formed sinks. This is necessary to construct a quantitative model for explaining this fragmentation. 
To perform this estimate,  we have selected all sinks at the age of 1000 yr within ten snapshots regularly spaced in time. 
Since we want to limit the influence that sinks may have on the gas dynamics, we select the objects that have at most one 
neighbour within 200 AU.  
We focus on three quantities that are of particular importance for the analytical model developed below, namely 
the mean density, the Mach number and the PDF of the density fluctuations. For each selected object, we have 
computed the mean density, radial velocity, rms velocity (to which the mean radial velocity has been subtracted)
in concentric spherical shells with logarithmic spacing. We have also  computed the PDF of the density fluctuations
(that is to say the density divided by the mean density within the shell)
within two spheres of radius 240 AU and 800 AU. We have then summed all the results to compute the mean values.  

Figure~\ref{run1_stat} displays the results. Top panels show the mean density (red curve) as well as the standard 
deviation from it (blue shaded area). The blue line is the singular isothermal sphere, $n _{\rm SIS}= \rho_{\rm SIS} / m_p=
 c_s^2 / (2 \pi G r^2) / m_p$.
The mean density is proportional to $r^{-2}$ and about 10 and 7 times  $n _{\rm SIS}$ for run1 and run2 respectively.

The Mach number is shown in middle panels. For run1 it goes from about 5 at 10 AU to 10 at 300 AU  while 
for run2 it is more on the order of 4-6. In both cases, there are considerable fluctuations, which go from 0 to 
about 2 times the mean value. As discussed below, such Mach number values are indeed expected from 
 energy equipartition. 

The bottom panels portray the PDF of the density fluctuations. Note that the natural logarithm is used to 
compare with the usual lognormal distribution \citep{HF12}. The upper and lower blue curves represent the fluctuations PDF  
within 800 and 240 AU respectively. The red curve is a lognormal distribution with a width 
$\sigma^2 = \ln(1 + b^2 {\mathcal M}^2) $  for ${\mathcal M}=6$ and $b=0.8$. It is not a fit but as can be seen, it 
matches reasonably well the distributions except for the high density powerlaw tails that are reminiscent of 
the gravitational infall that is taking place. 
For run1, these values are entirely reasonable since the Mach number goes from 6 to 10
(the $b$ parameter is obviously degenerated and so $b=0.6$ with ${\mathcal M}=8$ is equivalent).
It is compatible with the idea that many compressible modes \citep{Federrath08} are generated by the infall. 
 Interestingly, the shapes of the four PDF are all similar and the value $b=0.8$ with ${\mathcal M}=6$
may sound a bit large for run2.  
This is likely 
a consequence of the gravo-turbulence that dominates in these collapsing regions. 
Indeed the powerlaw tails, which are of gravitational origin, indicate that the fluctuations
with respect to the mean density (that we recall is itself proportional to $\rho_{\rm SIS}$)
are themselves collapsing.

\setlength{\unitlength}{1cm}
\begin{figure*}
\begin{picture} (0,19)
\put(0,12){\includegraphics[width=8.7cm]{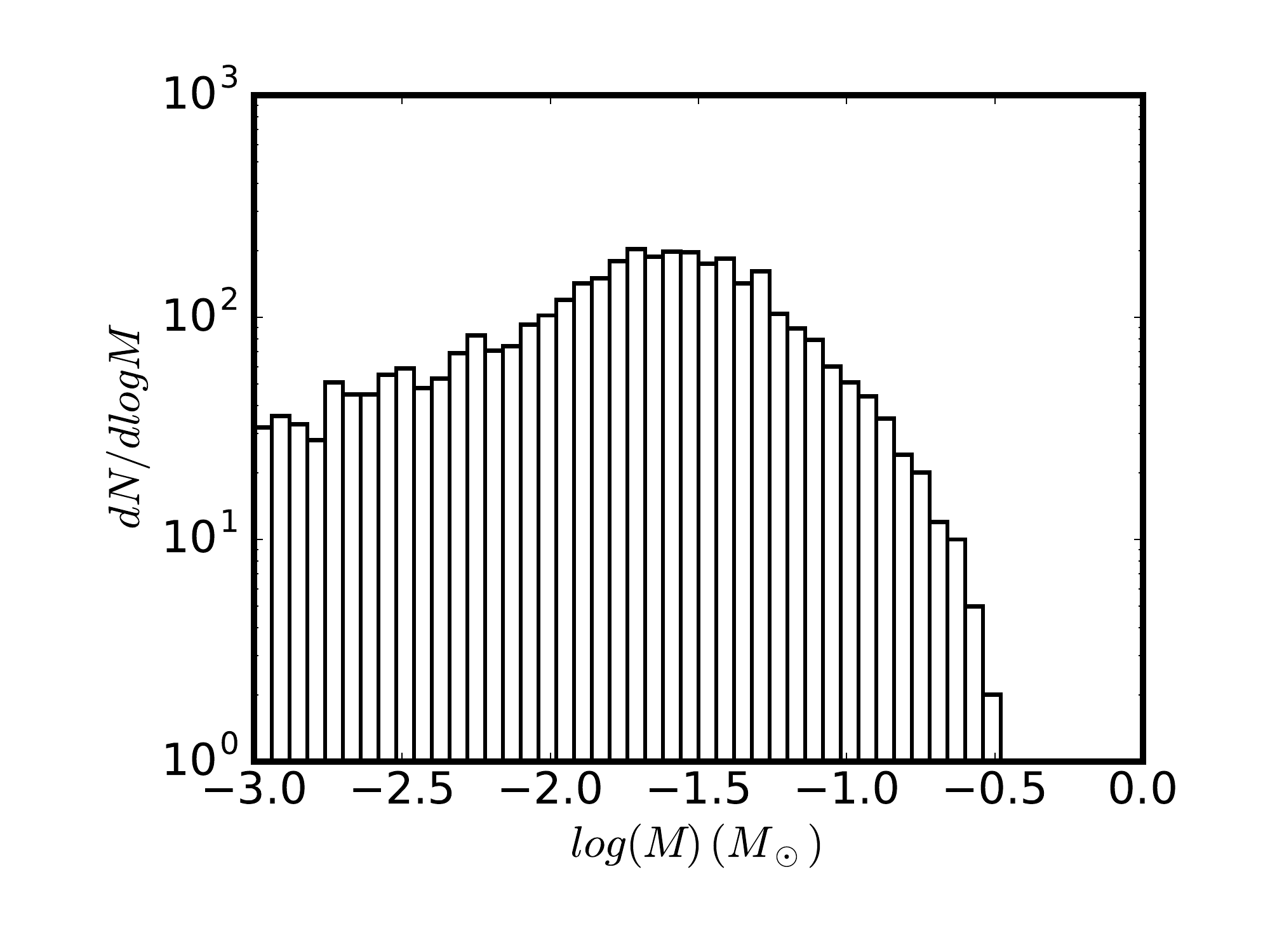}}  
\put(9,12){\includegraphics[width=8.7cm]{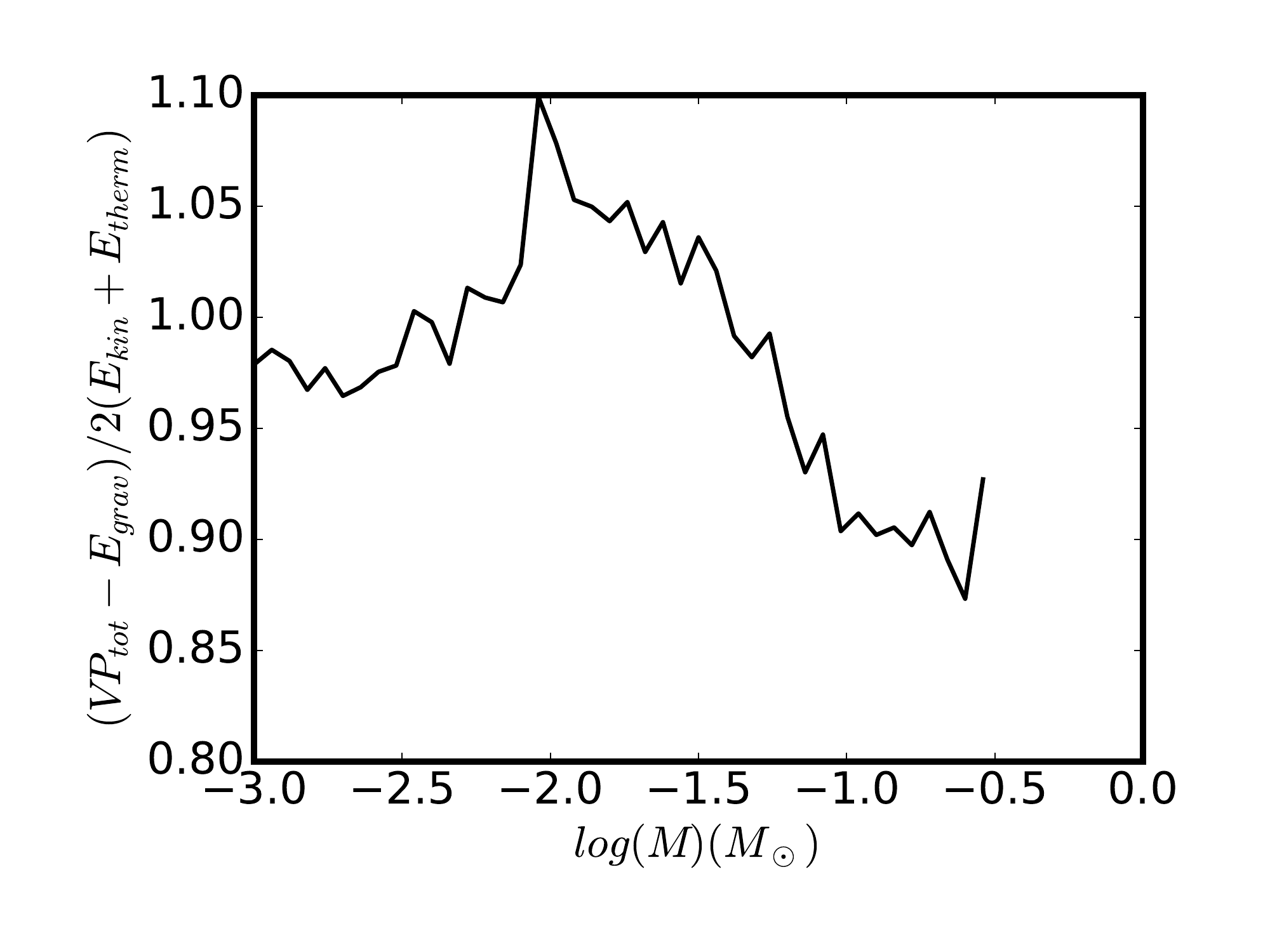}}  
\put(9,6){\includegraphics[width=8.7cm]{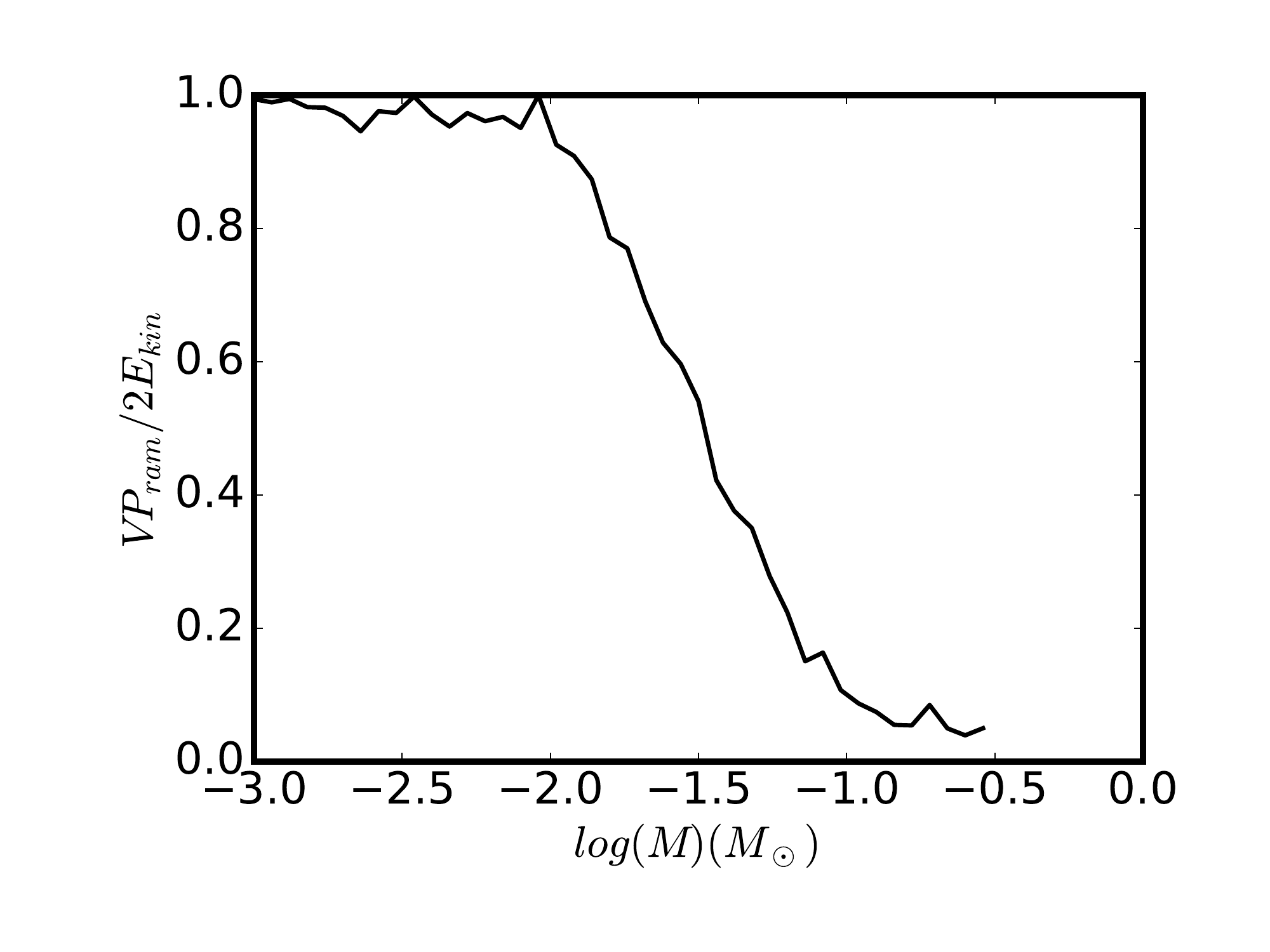}}  
\put(0,6){\includegraphics[width=8.7cm]{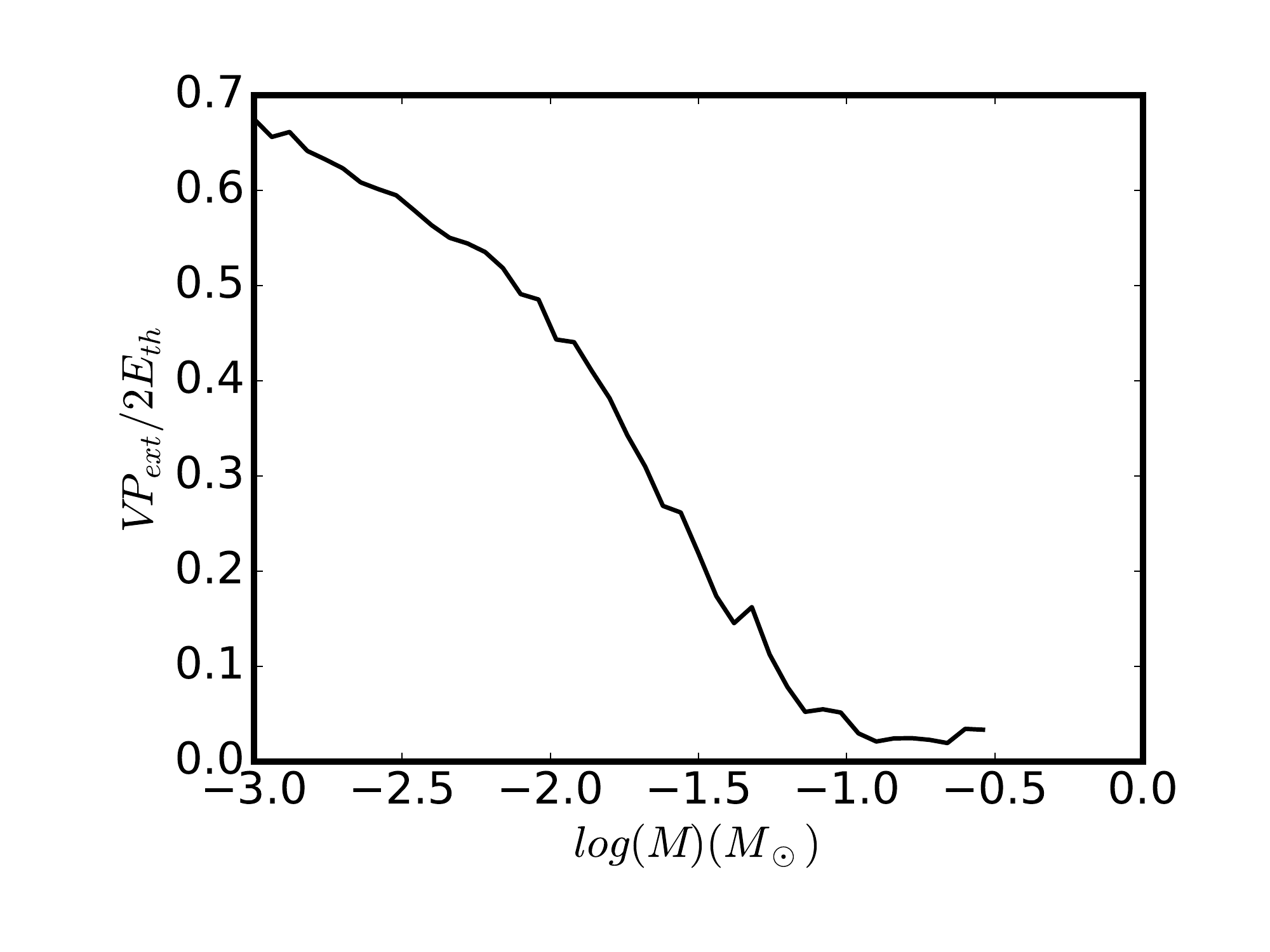}}  
\put(9,0){\includegraphics[width=8.7cm]{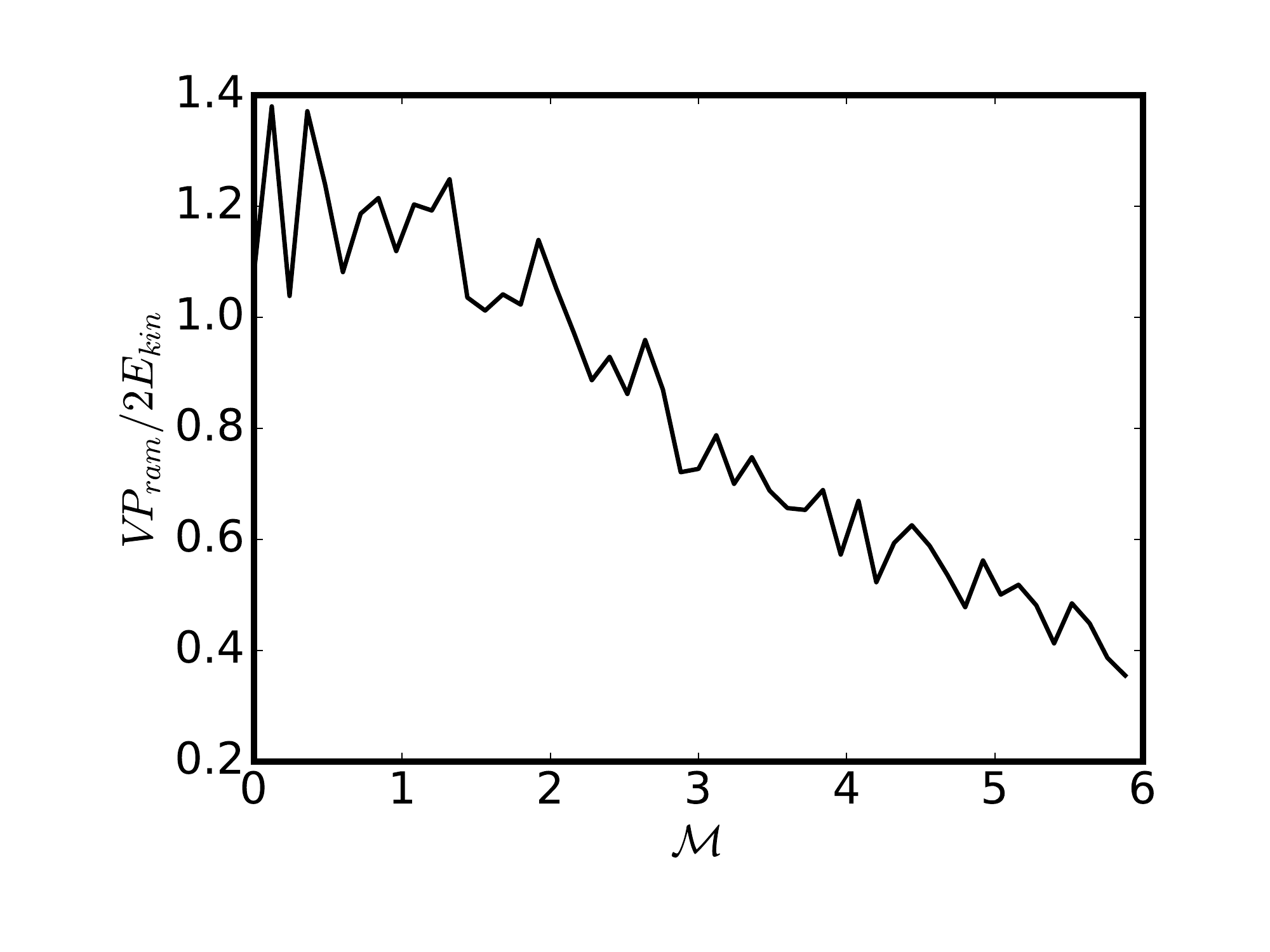}}  
\put(0,0){\includegraphics[width=8.7cm]{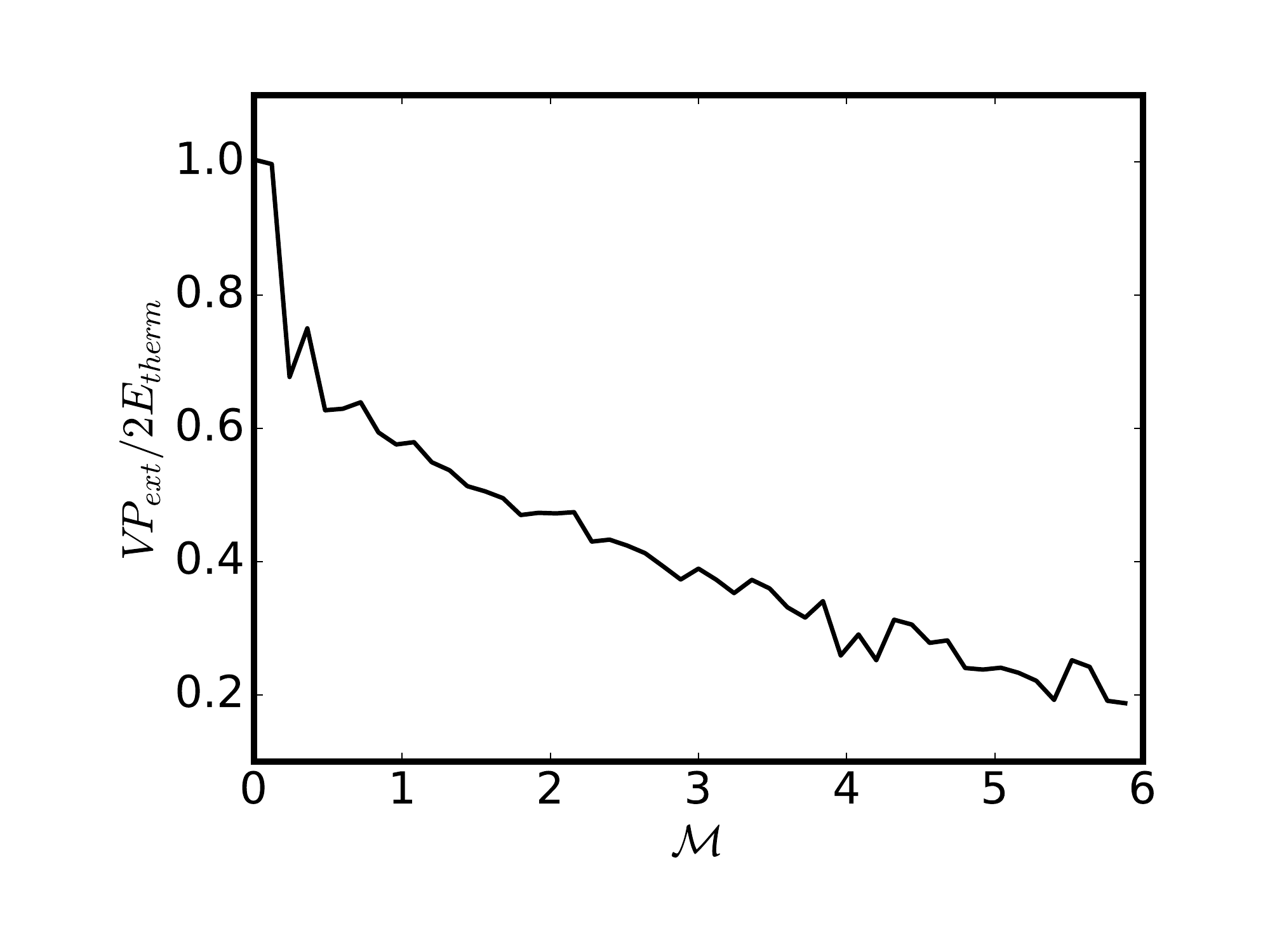}}  
\end{picture}
\caption{{\it Core} analysis and statistics for run 1. Top left panel displays the {\it core} mass spectrum.
Top right panel shows the radio of confining over support terms which appear in the virial theorem showing
that {\it cores} are remarkably virialised. Middle left and middle right panels respectively 
portray the ratio of thermal 
and kinematic surface of bulk terms. As can be seen surface terms while negligible for {\it cores} of mass
larger than $0.1 ~M_\odot$ are comparable to the bulk values for cores of mass below $M_L$. 
Bottom right and middle panels show the same quantities but plotted as a function 
of the Mach numbers instead.}
\label{run1_clump}
\end{figure*}

\subsection{{\it Core} statistics}
\label{core}
We now turn to the analysis of the self-gravitating gas structures that lead to the formation of stars and 
that we name {\it cores} although we warn that the structures we identify may be too dense and also 
not sufficiently isolated to correspond to most of the observed cores \citep{ward2007}. The underlying 
picture is however the same, i.e. coherent gas structure that are dominated by their gravity. 
The motivation to perform such analysis is to understand how fragmentation occurs in the vicinity 
of protostars, we need to understand how self-gravitating perturbations develop. In particular as our analytical 
model below uses the virial theorem, we want to verify that the {\it cores} are virialised but also 
we want to estimate the importance of surface terms which appear in the virial theorem.

To identify the {\it cores}, we proceed as in \citet{h2018} and \citet{nh2019}, we use the 
HOP algorithm \citep{eh1998} with a density threshold of 10$^8$ cm$^{-3}$. As the 
algorithm captures many density fluctuations that are close to the density threshold, we keep
only those for  which the gravitational energy dominates over the thermal one. 
Figure~\ref{run1_clump} presents various statistics inferred from  this sample. 

Top panel displays the core mass spectrum. Interestingly, it peaks at about $1-2 \times 10^{-2}$ M$_\odot$, 
i.e. close to  $M_L$.  At large masses, it drops relatively steeply. This is therefore in good agreement with the
idea that {\it cores} with a mass close to $M_L$ form an object (a sink in our case) with a mass that is about 
$M_L$ and then further keep accreting the surrounding gas until the coherence with it is lost. 

Before estimating the various contributions of the virial theorem, it is worth reminding that a virialised core 
is such that
\begin{eqnarray} \label{eq_vir} 
 E_\mathrm{g} + 2E_\mathrm{th} 
+ 2E_\mathrm{kin} -  {\rm VP}_{ext} - {\rm VP}_{ram} =0,
\end{eqnarray} 
where the various term are the gravitational energy, the thermal and the kinetic ones while 
${\rm VP}_{ext}$ and  ${\rm VP}_{ram}$ stand for the integrated 
thermal and ram pressure in the outskirt of the {\it cores}.
As detailed in section~\ref{vir_expr},  their exact expressions are
\begin{eqnarray} \label{ptotv} 
{\rm VP}_{ext}  &=  \oint P_{th} \, {\bf r \cdot dS}, \\
{\rm VP}_{ram}  &=  \oint \rho \, ({\bf v \cdot r})  ({\bf v \cdot dS}).
\nonumber
\end{eqnarray} 
The exact values of these two terms is not straightforward to estimate. Dimensionally they are similar to the thermal and 
kinematic energy respectively but their exact values require a surface integration. Particularly important are the
ratios $V P_{ext} / 2 E_{th}$ and $V P_{ram} / 2 E_{kin}$. As discussed in section~\ref{vir_expr}, it is expected that 
for small cores the contrast between the bulk and the surface thermal and kinetic energies is likely small while it 
becomes large for massive cores.  \citet{dib2007} have estimated the importance of the virial surface terms in the context of 
collapsing cores and found it to be comparable with the volume terms.  
Top right
 panel of Fig.~\ref{run1_clump} shows the mean value of $(-E_{grav} + V P_{tot}) / (2 E_{kin} + 2 E_{th}) $
in logarithmic bins of mass. The selected {\it cores} appear to be remarkably virialised, therefore 
stressing the importance 
of the surface terms.  It is therefore important to estimate them quantitatively to develop 
analytical models.

Middle panels of Fig.~\ref{run1_clump} portray the mean values of 
$V P_{ext} / 2 E_{th}$ and $V P_{ram} / 2 E_{kin}$
 per bin of core mass. 
It is found that both of them are almost 0 for $M \ge 0.1~ M_\odot$.
For $M < 10^{-2} ~M_\odot \simeq M_L$ it is found that $V P_{ext} / 2 E_{th} \ge  0.5$ while 
$V P_{ram} \simeq 2 E_{kin}$. Between these two regimes, both ratios decrease linearly with $\log M$.
As discussed in section~\ref{vir_expr} however, it is likely the case that the Mach number within the cores is more relevant 
to infer the dependence of their ratios. Therefore bottom panels display the same quantities as a function 
of the Mach numbers. We see that roughly linear dependences are inferred. 
At small Mach numbers, the ratios are the highest with values near unity while for large Mach numbers,
the surface terms become negligible.  Below expressions inferred 
from these plots will be used, namely
\begin{eqnarray} \label{VP_fit} 
{\rm VP}_{ext}  & \simeq   E_{ th} \times (0.6 - 0.1 {\mathcal M}), \\
{\rm VP}_{ram}  & \simeq    E_{ kin} \times (1.3 - 0.18 {\mathcal M}).
\nonumber
\end{eqnarray}

\section{Relation between the mass of the first hydrostatic core and the mass of the stars: analytical model}
\label{sec_analyt}


\subsection{Model assumptions and definitions}
In this paper, we attempt to construct a ``minimum'' model that characterizes  
the essence of the physics responsible for determining the {\it peak} of the IMF.
In this model, star formation proceeds chronologically throughout the following fundamental stages:
\begin{itemize}
\item[1)] at the very early stages of star formation, large-scale turbulence  generates overdense fluctuations
		with a power law mass spectrum. Some of these fluctuations, which
		can be seen as initial {\it mass reservoirs}, become
	gravitationally unstable,
decoupling themselves from the surrounding low density medium. \\
\item[ 2)] 
in order to lead eventually to the formation
of a protostar, the above density fluctuations must have a {\it minimum  mass} equal to the mass of the FHSC, $M_L$, which is typically of the order of a few 10$^{-2}~M_\odot$ \citep{larson69,Masunaga98,vaytet2013,Vaytet17}. 
As discussed in section~\ref{larson_core}, 
the main reason is that cores with $M<M_L$ are not massive enough to trigger the second collapse without 
radiating away a large fraction of their thermal energy. However it takes 
a long time for objects of mass below $M_L$ to cool (see the discussion in  section~\ref{larson_core}). \\
\item[ 3)] within the envelope, turbulence-induced density fluctuations are likely to be generated on top of the average $r^{-2}$ profile,  
leading to the formation of new collapsing fragments, within the same reservoir, providing their own mass is also at least $M_L$.
These new objects, which we call {\it shielding} fragments, accrete the gas in their 
surrounding and thus prevent further gas accretion onto the central object.  
Another effect which is likely playing a role in protocluster is that the new formed object
will act differently on the central object and its gas reservoir because the latter dissipates 
kinetic energy but not the former.  This would push the star away from its 
initial mass reservoir
and certainly reduces, or even possibly stops in the most extreme case, 
the accretion onto the central object. Appendix~\ref{sink_vel} explores this effect and suggests
that this indeed is happening. \\
\item[4)] the total number of such fragments within the collapsing envelope is thus crucial in setting up the final mass of the central object. It is where tidal forces induced by this latter and the surrounding envelope
play a crucial role by shearing apart density fluctuations. 
Therefore, at a given distance $r$ from the central core, only density fluctuations above a certain threshold accounting for this tidal shear contribution can become unstable and lead to the formation of a nearby core of mass $\ge M_L$. \\
\item[5)] 
Therefore, the final mass of a collapsing core is about the mass contained within a typical accretion radius $r_{\rm acc}$ in the envelope, which is the radius containing the characteristic number $n_f$ of unstable fluctuations necessary to prevent 
accretion of gas located further away than $r_{\rm acc}$.   
\end{itemize}
Whereas the first item is the basic building block of the gravoturbulent
theories \citep{padoan1997,HC08,hopkins2012}, steps 2-5 highlight the importance of
further fragmentation within the collapsing prestellar core envelope and the key role
played by tidal interactions in limiting the final mass of the central object,
as revealed by recent numerical simulations of a collapsing 1000 $\msol$ clump (paper II).

Quantifying precisely this series of processes is difficult as they are all complex 
and non-linear. The number of fragments, $n_f$, necessary to halt accretion for $r \ge r_{acc}$, in particular,
is not straightforward to estimate, as 
it is a highly non-linear phenomenon, which involves N-body interactions. 
We thus start with a set of simplifying approximations, among which
we assume that $n_f$ is typically equal to 2-5 as suggested by the numerical simulation results.  
In reality as shown in Fig.~\ref{run1_neigh}, there is  
a distribution of $n_f$ values, which may vary from place to place with the local conditions. 
However, as will be seen later, the exact value of $n_f$ has a modest influence on the result.

\subsection{Mathematical description}

\subsubsection{The mean envelope density}
The density distribution in the collapsing envelope around the accreting 
object is assumed to follow the one of a collapsing sphere \citep{larson69,Shu77,chang2015}: 
\begin{eqnarray}
\rho _{env}(r) &= {A c_\mathrm{s}^2 \over 2\pi G} {1 \over r^2},
\label{rho_env}
\end{eqnarray}
where $A$ defines the amplitude of the density profil  and $c_s$ denotes the speed of sound.
As shown above (see Fig.~\ref{run1_stat} and Fig.~4 of paper I), numerical simulations suggest 
$A\sim 5-10$. 
For convenience we define $r_{e,L}$  as  $M_L= 2
A ( c_\mathrm{s}^2 /  G)  r_{e,L}$. That is to say $r_{e,L}$ is the radius for which 
the mass within the envelope is equal to $M_L$. We stress that it is not the radius 
of the Larson core itself which is typically on the order of 5-10 AU. For typical values,
$A=10$ leads to $r_{e,L} \simeq 20 AU$.
This yields:

\begin{eqnarray}
M_{env}  (r) &= \int \rho _{env}(r) 4 \pi r^2 dr=2 A { c_\mathrm{s}^2 \over G}  r = M_L \widetilde{r},
\label{m_env}
\end{eqnarray}
where $\widetilde{r}=r/r_{e,L}$.

\subsubsection{The density fluctuations within the envelope}
\noindent As mentioned above, the envelope is dominated by turbulent motions,
leading to density fluctuations $\delta_\rho$ on top of the mean density $\rho _{env}(r)$, 
given, in reasonable approximation, by a lognormal distribution
\begin{eqnarray}
\label{lognorm}
P(\widetilde{r}, \delta_\rho(\widetilde{r})) = {1\over \sqrt{2 \pi} \sigma(\widetilde{r})} \exp \left[{-\left(\delta_\rho(\widetilde{r})+\sigma^2(\widetilde{r})/2\right)^2 \over 2\sigma^2(\widetilde{r})}\right],
\end{eqnarray}
with $\delta_\rho(\widetilde{r}) = \ln (\rho(\widetilde{r}) / \rho _{env} (\widetilde{r})) = \ln (1+\eta(\widetilde{r}))$,
with $\eta(\widetilde{r})=\rho(\widetilde{r})/\rho _{env}(\widetilde{r}) - 1$ and $\sigma$ defines the width of the density PDF.
The density fluctuation is related to the {\it local} turbulent Mach number as
$\sigma^2(\widetilde{r}) = \ln \left(1+b^2\mathcal{M}(\widetilde{r})^2\right)$,  \citep[e.g.][]{HF12}. 
To estimate the local Mach number $\mathcal{M}(\widetilde{r})$, 
we assume that the energy in turbulent motions is a fraction $\epsilon\le 1$ of the gravitational energy:
\begin{eqnarray}
\label{eq_mach_2}
\mathcal{M}(\widetilde{r}) 
&= {1 \over c_\sound} \sqrt{\epsilon{G [M_{\rm L} + M _{env}(r)] \over r} }
= \sqrt{2 A (1+{1 \over \widetilde{r}}) \epsilon}. 
\end{eqnarray} 
As seen from Fig.~\ref{run1_stat}, which displays the  PDF  of the density fluctuations, $\delta \rho / \rho$, around sink particles, 
the lognormal PDF with $b=0.8$ and $\mathcal{M} \simeq 6$ is a good approximation. This corresponds to $\epsilon \simeq 1$. 
Note that the Mach number distribution seen in Fig.~\ref{run1_stat} shows a tendency for $\mathcal M$ to increase 
with $r$ instead of decreasing. This is obviously a consequence of the large scale environment that is not accounted
for in our simple estimate.

 Let us stress that the typical scales discussed here are  a few hundred AU's, which are 
difficult to probe observationally because of the lack or reliable tracers. 
Supersonic collapsing motions are clearly observed toward 
the center of at least some cores \citep{ward2007,difra2007} and it may be difficult to 
separate the infall from the turbulence. Also the regions under investigation 
correspond to massive star forming clumps \citep{lee2016a,traficante2018} rather than low mass cores.


      \begin{table*}
         \begin{center}
            \begin{tabular}{lccc}
               \hline\hline
               Name & Nature  & Typical value  \\
               \hline
               $M_L$ &  Mass of the first Larson core   & 0.02-0.03 $M_\odot$     \\
               A &  Density of the envelope   & 5-10      \\
               $\epsilon $  & Density fluctuations and Mach number    & 0.3-1.      \\
               $\alpha_{turb}$  & Turbulent support  & 0-1    \\
               $n_f$  & Number of shielding fragments  & 2-5  \\
               \hline
            \end{tabular}
         \end{center}
         \caption{Summary of the model parameters.}
\label{table_param}
      \end{table*}

\subsubsection{The instability threshold}
\label{insta_thres}
 As discussed above and calculated in paper II, 
only density fluctuations exceeding a certain threshold, $\eta_{crit} (\widetilde{r})$, eventually collapse.
This is because the self-gravity of the perturbation must supersede the tidal forces and the local 
 thermal and turbulent support. Writing the virial theorem for a density perturbation, 
located at $r\p$, of size $\delta r\p$ and of overdensity $\eta_p = \rho_p / \rho _e -1$
we get
\begin{eqnarray} \label{eq_int_vir} 
E_\mathrm{vir}&(r\p,\delta r\p,\eta _p) = E_\mathrm{g}(r\p,\delta r\p,\eta_p) + 2E_\mathrm{ther} 
+ 2E_\mathrm{kin} - {\rm VP}_{ext} - {\rm VP}_{ram}    
\end{eqnarray} 
In this expression, the gravitational energy is given by 
\begin{eqnarray} \label{Egrav} 
E_\mathrm{g}(r\p,\delta r\p,\eta_p) = \int\limits_{V\p} (\rho\e+\rho\p)~(\mathbi{g}_\mathrm{L}+\mathbi{g}\e+\mathbi{g}\p) \cdot \delta\mathbi{r} dV,
\end{eqnarray} 
where $\rho\e$ and $\rho\p$ are the density of the envelope and the perturbation, respectively, and 
$\mathbi{g}_\mathrm{L}$, $\mathbi{g}\e$, $\mathbi{g}\p$ the gravitational fields due to 
the central Larson core, the envelope and the perturbation itself. Their expressions are given 
in paper II. 

To compute the thermal energy, it is necessary to know the temperature. 
For the sake of simplicity, the calculations presented 
below assume that the gas remains isothermal. This assumption is discussed in section~\ref{env_temp}.

The kinetic energy in Eq.~(\ref{eq_int_vir}) requires the knowledge of the velocity dispersion within the perturbation. 
We assume that usual scaling laws also apply here so that 
the velocity dispersion, $\delta v$, within
 a fluctuation of size $\delta r$ is proportional to $\delta r^{0.5}$. Thus
\begin{eqnarray}
E_\mathrm{kin} = 0.5 M_p \delta v ^2 \simeq 0.5 \alpha_{turb}  M_p c_s^2 {\cal M}^2(r) \times {\delta r \over r},
\label{Ekin}
\end{eqnarray}
 where $M_p$ is the mass of the perturbation and $\alpha_{turb}$ is a coefficient of the order of a few. 

Finally, the impact of the external confining pressure at the perturbation boundary must also be considered in Eq.~(\ref{eq_int_vir}). 
As discussed above (see Fig.~\ref{run1_clump}) and in the appendix,
 estimating this quantity is difficult. The numerical estimate shown in Fig.~\ref{run1_clump} suggests that 
in the range  $M_L < M < 10 M_L$, we have 
the expressions stated by Eqs.~(\ref{VP_fit}) .

\setlength{\unitlength}{1cm}
\begin{figure}
\begin{picture} (0,9)
\put(0,4.5){\includegraphics[width=8.7cm]{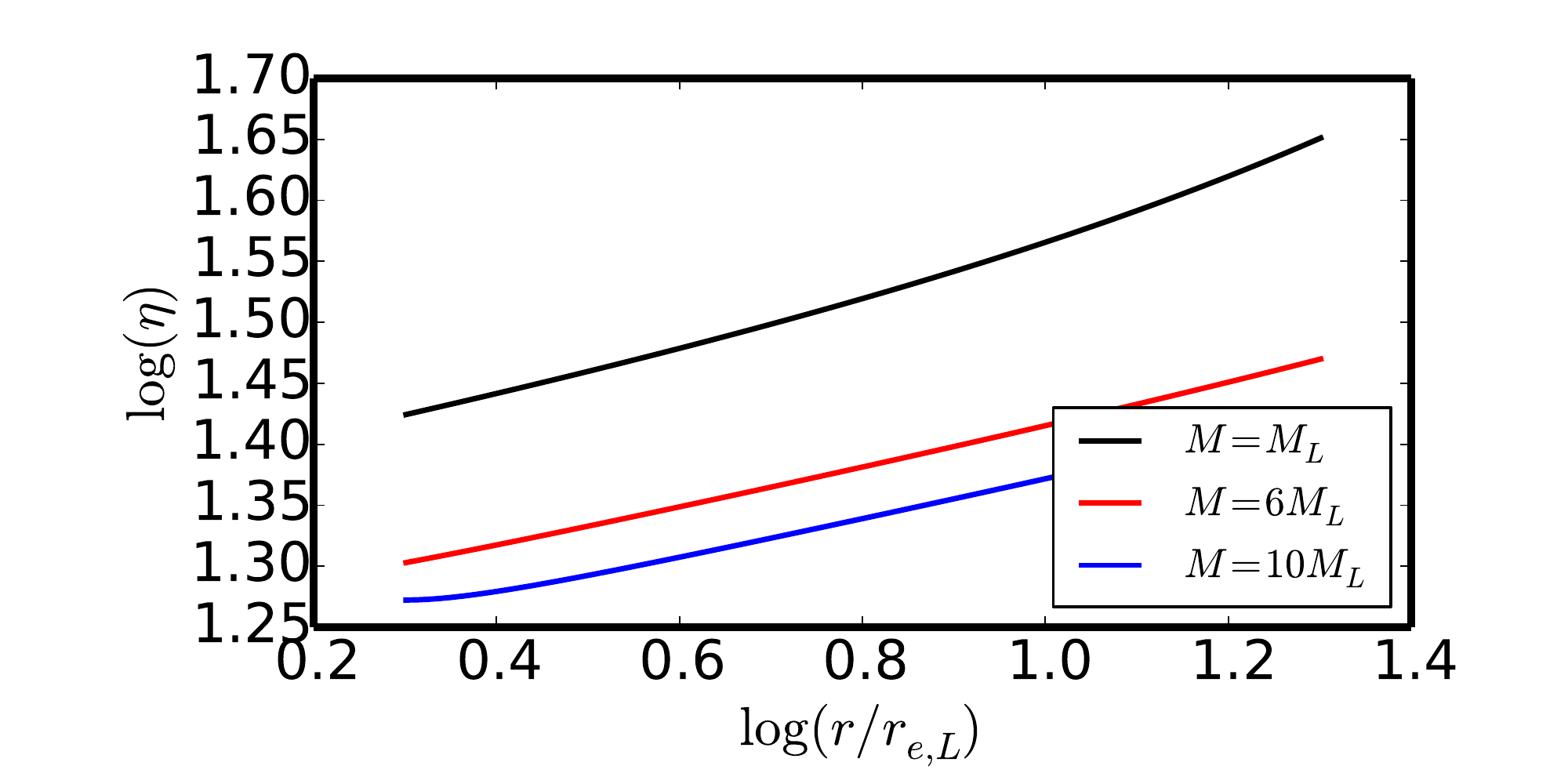}}  
\put(0,0){\includegraphics[width=8.7cm]{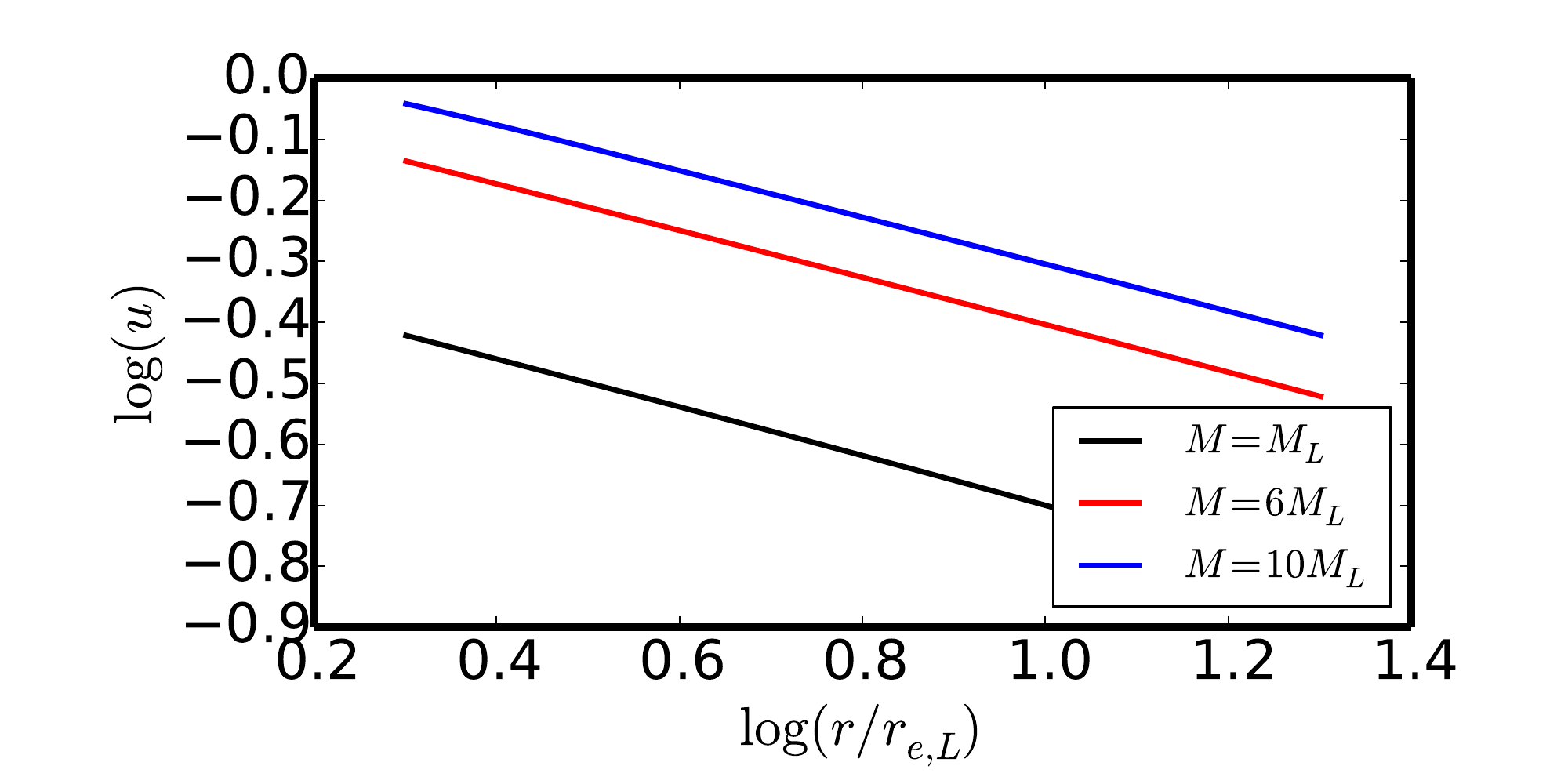}}  
\end{picture}
\caption{Critical density fluctuation amplitude, $\eta_{crit} = \exp( \delta_{p,crit}) -1 $ (top panel) 
and relative perturbation radius, $u=\delta r_p / r_p$ (bottom panel) for 
run1 case (i.e. $A=10$, $\mathcal{M}=8$) and various fragment masses. The surface terms
(see Eqs.~\ref{ptotv}) are given by Eqs.~(\ref{VP_fit}). Typically, only density fluctuations
20 to 40 times above the local mean density are unstable. }
\label{eta_u_r}
\end{figure}

Top panel of Fig.~\ref{eta_u_r}   displays the critical density fluctuation, 
$\eta_{crit}(\widetilde{r}) = \exp( \delta_{p,crit}) -1$,
and the perturbation relative size, i.e. $u = \delta r_p / r_p$
for various mass of the perturbation. The model parameters correspond to the ones measured 
for run1, i.e. $A \simeq 10$ and ${\mathcal M} \simeq 8$ (see Fig.~\ref{run1_stat})
As can be seen typical unstable density fluctuations are about 20-40 times 
the local mean density. This high value stems for the high turbulence that is adopted here.
As expected bigger masses require density fluctuations that are less intense but bigger in size
than smaller ones.

\subsubsection{Mean number of self-gravitating fluctuations of mass $\ge  M_L$}

\setlength{\unitlength}{1cm}
\begin{figure}
\begin{picture} (0,9)
\put(0,4.5){\includegraphics[width=8.7cm]{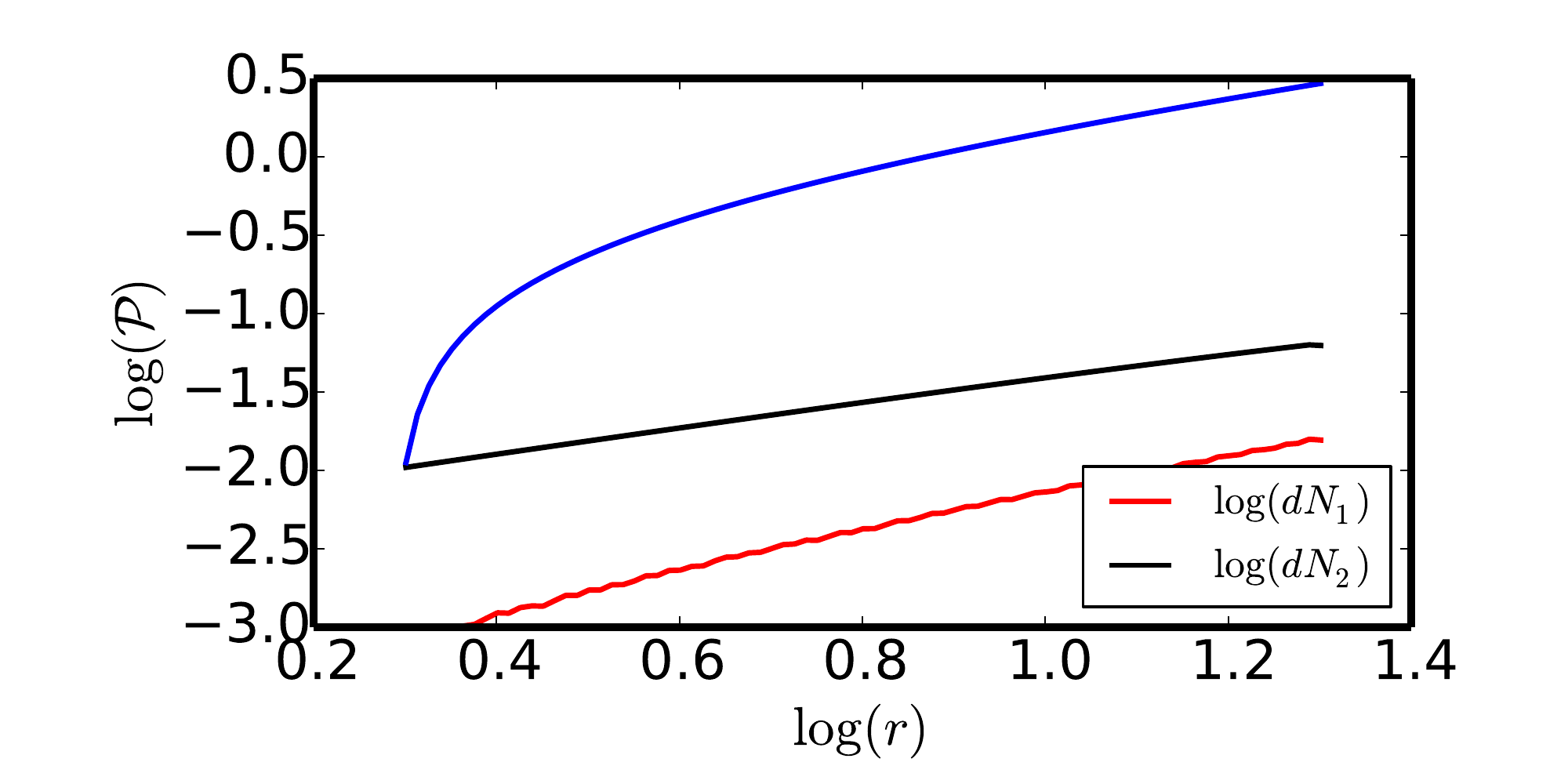}}  
\put(0,0){\includegraphics[width=8.7cm]{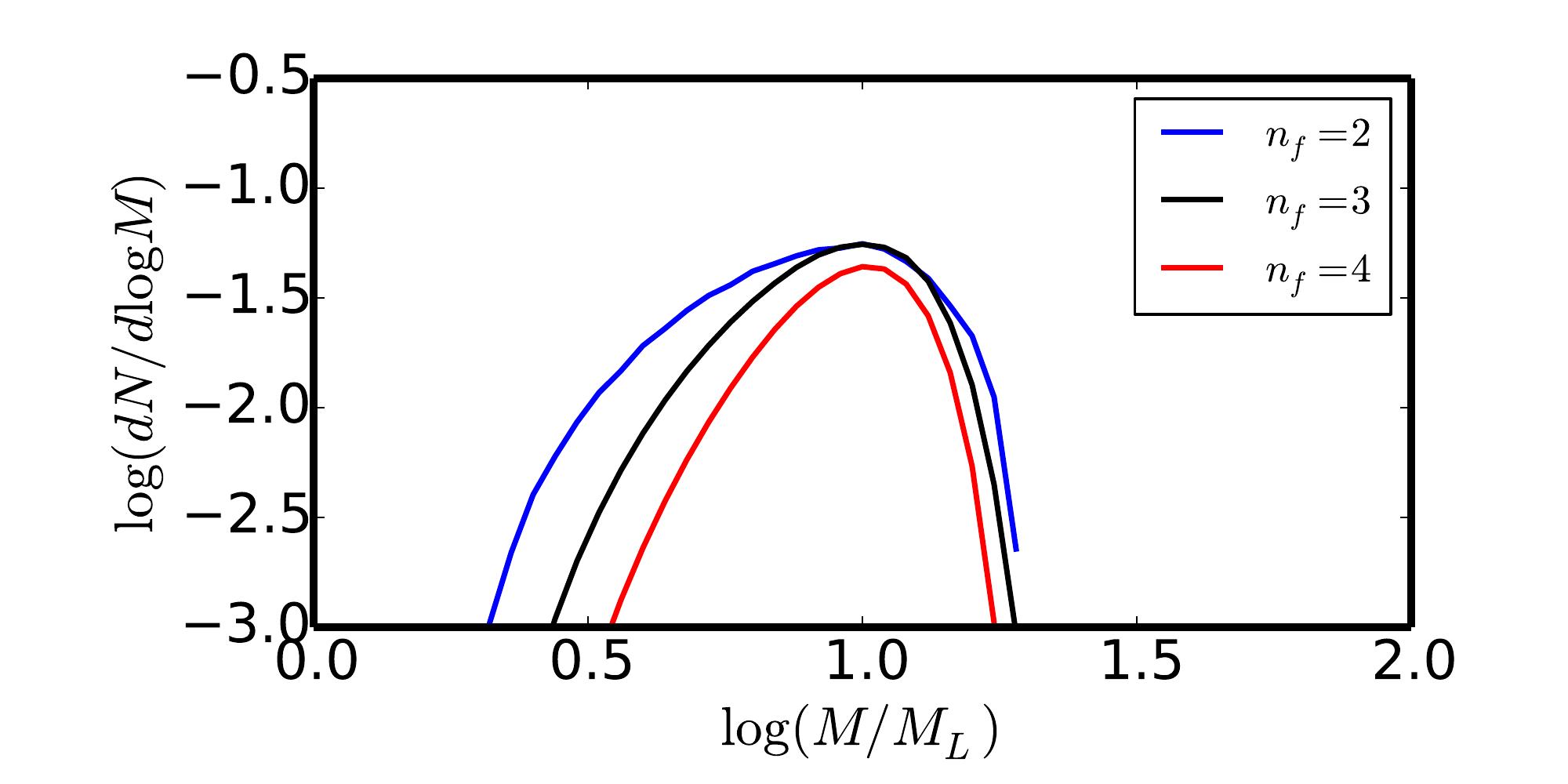}}  
\end{picture}
\caption{Top panel: mean number of fragments as a function of distance.
Blue curve: mean number of fragments (see Eq.~\ref{Ntot}) as a function of $\widetilde{r} = r / r_{e,L}$ for 
run1 case (i.e. $A=10$, $\mathcal{M}=6$). Red and black are the mean number of type 1 and type 2 fragments respectively 
(see Eqs.~\ref{dN1} and~\ref{dN2}. )
in logarithmic shells. The mean number of fragments is equal to one at about $r \simeq 9 r_{e,L}$.
Bottom panel: mass distribution of $M_{unsh}$ (see Eq.~(\ref{unshield})) for $n_f=$2, 3 and 4  shielding fragments.
The distribution peaks at typically 10 $M_L$.}
\label{mean_frag}
\end{figure}

We now estimate the mean number of self-gravitating fluctuations of mass 
at least equal to $M_L$ as a function of the distance, $r$, from the central fragment. 
Since the density threshold $\delta_c$ varies with the 
radial distance $r$, the probability to find unstable fluctuations must 
be performed in concentric shells inside which this probability is uniform.
The first step is to estimate the number of relevant fluctuations within the shells.
Two types of fluctuations must be considered.

 First, we have the ones 
whose mass is larger than $M_L$ while their density, $\eta_{p,crit}$ and radius, $\delta r _p$, are 
the results of the virial equilibrium expressed by Eq.~(\ref{eq_int_vir}). As the 
volume they occupy is: $\delta V= 4 \pi / 3 \delta r _p ^3$, a shell of radius $r$ and 
thickness $dr$ contains up to $3 (r / \delta r_p)^3 dr / r$ perturbations of this size. 
The number of unstable fluctuations in the shell is equal to the the probability to find a density 
fluctuation in the desire range of density times the number of fluctuations of this size.
 Within the shell the mean number of self-gravitating fluctuations 
of mass $\ge M_L$ is therefore given by
\begin{eqnarray}
\label{dN1}
dN _1(r) &= \int _ {\delta_{min}} ^ {\delta_{max}}  
{3 \over \sqrt{2\pi} \sigma_0 } \exp \left[{-\left(\delta+\sigma_0^2/2\right)^2 \over 2\sigma_0 ^2}\right]
\times  \left( { r \over \delta r _p } \right)^3 { dr \over r}
 d \delta.
\end{eqnarray}
where $\delta_{max}= \delta_{p, crit}$ is the density corresponding to a fluctuation of mass $M_L$ 
(Fig.\ref{eta_u_r}) and 
$\delta_{min}$ the density for which $\delta r _p / r = 0.5$, i.e. the size of the fluctuation 
is half the radius $r$ that we adopt as an upper threshold . We have tested other values and found 
that it does not drastically affect the result. 

The second type of fluctuations that must be considered are the ones whose mass is equal to 
$M_L$ but the density is larger than $\delta_{p, crit}$. The size of these fluctuations 
is given by $\delta r _{p,crit} \times \left( {\rho_{p,crit} \over \rho_p} \right)^{1/3}$. 
The density term takes into account the fact that denser fluctuations have a smaller size 
since we integrate over fluctuations of mass $M_L$.
The number of self-gravitating fluctuations with mass equal to $M_L$ is thus
\begin{eqnarray}
\label{dN2}
dN _2(r) &= \int _ {\delta_{p,crit}} ^ {\infty}
{3 \over \sqrt{2\pi} \sigma_0 } \exp \left[{-\left(\delta+\sigma_0^2/2\right)^2 \over 2\sigma_0 ^2}\right]
\times  \left( { r \over \delta r _{p,crit} } \right)^3 {\rho _p \over  \rho _{p,crit}  }  { dr \over r}
 d \delta.
\end{eqnarray}

Within a sphere of radius, r, the number of self-gravitating fluctuations of mass larger or 
equal to $M_L$ is thus 
\begin{eqnarray}
\label{Ntot}
N(r) = \int _0 ^r (dN_1 + dN_2).
\end{eqnarray}

\subsubsection{Mean mass calculation}
To obtain the  mass distribution, we proceed as follows.

The probability to find $n_f$ fragments inside the sphere of radius $r_i=r_{acc}$ with 
at least a fragment in the shell $i$ is 
\begin{eqnarray}
P(r_{acc},n_f) = \sum _{n_i=1, n_f ; \sum n_j = n_f} \prod _{j=0,i} p_j ^{n_j}.
\label{big_proba}
\end{eqnarray}
where $p_j^{n_j}$ is the probability to find $n_j$ fragments in shell $j$. 
When $d N$ is small, we simply have $p_j^1 \simeq d N(r_j)$ and 
$p_j^0 \simeq 1 - p_j^1$.
The product in Eq.~(\ref{big_proba}) runs over all shells inside (including) shell $i$, the 
total number of fragments being $\sum n_i = n_f$, and the sum is over 
all the fragment distribution. For example the probability to find 2 fragments inside 
$r_3$ is given by 
$P(r_2,3)= p_3^2 p_2^0 p_1^0 + p_3^1 p_2^1 p_1^0 + p_3^1 p_2^0 p_1^1 $. Note that as we use a large number of 
shells (typically 90-100, logarithmically spaced in $r$), the shells are all thin and the probability
to have more than one fragment within the same shell is negligible. We have verified that our result does not 
depend on the shell number.

The mass, $M_{acc}=M_{env}(r_{acc})$, inside radius $r_i=r_{acc}$ is  given by Eq.~(\ref{m_env}). In combination 
with Eq.~(\ref{big_proba}), this gives the overall mass distribution, $dN_{acc} / dM = P_{acc}$. 
All of the  mass,  $M_{acc}$,  however, is not going to be accreted because some
of the fragments lie in 
shells contained within  shell $i$ and thus can accrete a fraction of $M_{acc}$. Therefore, we  compute the mass that is  
located at radius smaller than  the fragment position, with the assumption that a fragment located in shell $j$ 
accretes a fraction $1/n_f$  of the mass located outside the sphere of radius  $r_j$. That is to say we assume that 
all fragments shield  $1/n_f$ of the infalling gas. 
  We call this mass (a fraction of $M_{acc}$), which has {\it not} been accreted
 by the other (shielding)
 fragments and then will eventually be accreted by the central seed the {\it unshielded} mass
$M_{unsh}$. It is straightforward to see  that 
\begin{eqnarray}
M_{unsh} = { \sum _{l=1,n_f} m_l \over n_f},
\label{unshield}
\end{eqnarray}
where $m_l = M_{env}(r_l)$.
Note that $M_{unsh}$ depends on the fragment distribution, not only on the position of the most distant fragment.
To get the mass distribution, $dN_{unsh}/dM = P_{unsh}$,  we have thus computed the histogram of
 $M_{unsh}$ weighted by the probability $P(r_{acc},n_f)$.  

Clearly this mass is indicative and in particular the assumption that all fragments shield a fraction 
 $1/n_f$ of the mass, while reasonable for large samples, is likely to undergo large statistical 
fluctuations. Let us remind that the goal here is to estimate the peak position, i.e. a preferred mass, 
and not the full mass spectrum.

\subsubsection{Summary of model parameters}
Table~\ref{table_param} summarizes  the parameters of the model. We stress that their values 
are consequences of the physics, notably the dynamics which develop
during the collapse. They are estimated through numerical simulations 
and are  fluctuating by a factor of a few.

\subsection{Results and comparison with simulations}
\label{comparison}
We focus on model values that are directly inspired from the values measured in run1. 
We adopt $A=10$, ${\mathcal M}=8$, $b=0.6$. For the external pressure, we adopt a fit 
taken from section~\ref{core}. Let us stress that those are typical values but 
that as seen from Fig.~\ref{run1_stat}, large deviations from these values have been inferred,
implying that in reality one should sum over a large number of configurations. 
Another limit is that Eqs.~(\ref{VP_fit}) are obtained for {\it cores} that are already formed and developed
while the fluctuations we are identifying are more core progenitors.

\subsubsection{Results for value parameters inferred from the simulations}
Top panel of Fig.~\ref{mean_frag} shows  $N$ (blue curve), $dN_1$ (red curve) and $d N_2$ (black curve) as a 
function of $\widetilde{r}$ for a series of logarithmic shells (see Eqs.~\ref{dN1},~\ref{dN2} and~\ref{Ntot} 
for definitions). As can be seen the fragments of 
type 2 dominate (from Fig.~\ref{eta_u_r} this is because they are smaller in size and therefore more numerous).
The mean number of fragments, $N$, becomes equal to 1 for $r / r_{e,L} \simeq 8-9$
Therefore qualitatively, assuming that all the mass within this radius will be accreted onto the central fragment, we 
see that the final mass of this latter is expected to be $\simeq 10 ~M_L$.

The corresponding distribution of the unshielded masses as stated by Eq.~(\ref{unshield}) 
is portrayed in bottom panel of Fig.~\ref{mean_frag}, where the mass distribution 
is shown for 2, 3 and 4 shielding fragments. As can be seen it peaks at about $10~M_L$ and 
weakly depends on the number of shielding fragments, $n_f$.

Note that the validity of the distribution 
is not expected to extend over the whole mass spectrum and is more likely limited in the region of the peak. 
In particular 
the  high masses are likely determined by the reservoir distribution as discussed in paper I.

\subsubsection{Comparison between model and simulation results}
The distribution of the unshielded mass, which peaks at about $10~M_L$, is in reasonable agreement with the peak 
of the distribution observed in run1 and run2 (see Fig.~\ref{run_IMF}). We remind that $M_L$ is about 0.02-0.03 $M_\odot$
with our numerical implementation, which implies that in run1 and run2 the ratio between $M_L$ and the observed 
peak position is about 5-10. 

Since the distribution is inherited from the mean fragment number (see Eq.~\ref{Ntot}), it is worth 
comparing it with the  simulation results, that is to say the mean number of objects in systems, $<n_{sys}>$
shown in bottom panels of Fig.~\ref{run1_neigh}.
In left panel (run1), the four black curves represent respectively (from top to bottom)
the values ${\mathcal M}=6$, 8, 10 and  10 but with the assumption ${\rm VP}_{ext} = {\rm VP}_{ram}=0$. 
In right panel (run2), the adopted values are 4, 6 and 8. For run2 we used $A=7$ instead of 10 for run2. 
The width of the density perturbation distribution is chosen to be the same for both runs. 
In both cases a reasonable  agreement is obtained for the highest values of the Mach  number explored. 
The model with ${\rm VP}_{ext} = {\rm VP}_{ram}=0$ leads to too few  fragments.

The most important disagreement is that the model predicts very few fragments below 100 AU (i.e. $<n_{sys}> =1$) 
while in the simulations
this number is about 0.3-1 (i.e. $<n_{sys}> \simeq 1.3-2$ depending on the time and the run). 
The most likely explanation is that in the simulations fragments have a tendancy to migrate approaching each 
other because they are gravitationally bound to each other. Indeed the typical density at 100 AU is about $10^9$ 
cm$^{-3}$ which corresponds to a freefall time of 10$^3$ yr. Since this time is short and the fragments
likely fall toward each other even before a sink particle is introduced, this effect certainly introduces
some differences between the simulations and the model. It may for instance lead to a somewhat smaller 
ratio between the peak of the stellar spectrum and $M_L$.


\section{Conclusion}
We have performed numerical simulations with the adaptive mesh refinement code RAMSES to investigate 
the origin of the peak in the stellar mass function, i.e. the characteristic mass of stars. 
We use two completely different setups and initial conditions to verify that the 
results are not due to a specific choice. We confirm that the peak is typically equal to about 5-10 times 
the mass of the FHSC. 

To verify that the factor 5-10 is due to further accretion that is eventually 
limited by the fragmentation of the collapsing cloud at a distance of typically 100-200 AU, we have performed
complementary numerical experiments in which the fragmentation is forbidden up to 140 and 280 AU from existing stars. 
We found that as expected, the peak shifts to higher masses. 
To quantify this fragmentation, we have computed the typical number of neighbours at various distances and ages of the 
sink particles. 
We have also measured several properties such as density, Mach number and PDF of density fluctuations in the vicinity of 
the sink particles. And finally we have extracted the cores that eventually lead to new objects and compute the various terms which 
enter in the virial theorem. 

We have then presented an analytical model to derive the position  of the 
peak of the stellar initial mass function. It is based on the idea that, as the collapse of a star-forming clump is stopped
when dust becomes opaque to its own radiation, matter piles up until at least the mass 
of the FHSC, $M_L$, is accumulated.  Since there is gas surrounding the FHSC, it grows by accretion and this
growth is very important, according to us, in setting the characteristic mass of stars. 
Due to the density fluctuations within the collapsing envelope surrounding this central core, however, new fragments form, accreting material around them and thus limiting the accretion onto the
central object. The induced tidal forces between core and envelope, however, shear apart these density fluctuations and thus, 
together with thermal and turbulent support, limit
further fragmentation in the envelope. 

As a result,
the  distribution of the final central accreted mass presents a peak around about 10 $M_L$, as indeed
inferred  in numerical simulations. 
A quantitative comparison between the mean number of neighbours as a function of distance from the stars measured
in the simulations and inferred from the analytical model, reveal good agreement suggesting that the analytical 
model captures the most important physical processes responsible of setting the peak of the stellar mass function.


\begin{acknowledgements}
We thank the anonymous referee for their constructive comments that have improved the paper.
This research has received funding from the European Research Council under the European
 Community's Seventh Framework Programme (FP7/2007-2013 Grant Agreement no. 306483).
\end{acknowledgements}

\bibliography{lars}{}
\bibliographystyle{aasjournal} 

\appendix




\section{The physics of the First hydrostatic core: orders of magnitude and trends}
\label{larson_core}
As the present theory relies on the first hydrostatic, or so-called first Larson core, 
we first start by discussing orders of magnitudes and trends. We are in particular interested
by getting simple expression for its mass and typical life time. While several studies 
have performed detailed calculations, simple expressions would be useful to infer simple 
dependence. 

\subsection{Physics of the First hydrostatic core: simplifying assumptions}
First, we assume that the core is virialised
\begin{eqnarray}
{3 \over 5} {G M_L m_p \over R} \simeq {\cal H} \times  3  k_B T,
\label{hydrostat}
\end{eqnarray}
where $m_p$ is the mass per particles, $M_L$ is mass of the Larson core, $T$ and $R$ its temperature and radius.
 ${\cal H}$ is a parameter of the order of a few, which reflects that  
the hydrostatic equilibrium equations should be solved instead of mere virial equilibrium. We do so below 
but it is very insightfull to get a simple and explicit expression valid within a factor of a few. 
It will be convenient to express $M_L$ as a function of the density, 
$M_L = 4 \pi / 3 \rho R^3, thus $
\begin{eqnarray}
M_L  \simeq  {\cal H}^{3/2} \times  5^{3/2} \left( {k_B T \over G m_p} \right) ^{3/2 }\left( 3 \over 4 \pi \rho  \right)^{1/2}, 
\label{hydrostat2}
\end{eqnarray}
The hydrostatic core ends when $H_2$ molecule starts dissociating and second  
collapse occurs at about $T=T_{dis} \simeq 1500$ K. 
Eq.~(\ref{hydrostat}) provides a relation between $M_L$ and $R$. 
To estimate, say the radius as a function of mass, we need to know at which 
density, $\rho_{dis}$, the second collapse starts or equivalently how 
the density depends on the temperature. To achieve this we need to know 
the entropy of the fluid. Using the well known fact (that we verify below) that 
the cooling time of the core is several times larger than its dynamical time, we can 
assume that above a density, $\rho_{ad}$, to be determined, the gas is essentially adiabatic. 
At the beginning of this adiabatic phase, the rotational levels of $H_2$ are not excited
and the adiabatic index is $\gamma=5/3$. Then when the rotational levels of $H_2$  
get excited $\gamma$ drops and  at about $T_{ex} \simeq 150$ K, we have $\gamma=7/5$. Thus
\begin{eqnarray}
{\rho_ {dis}} \simeq \rho_{ex} \left( { T_{dis} \over T_{ex}}  \right)^{5/2}
\simeq  \rho_{ad} \left( { T_{ex} \over T_0}  \right)^{3/2} \left( { T_{dis} \over T_{ex}}  \right)^{5/2}.
\label{rho_T}
\end{eqnarray}
where $T_0$ is the mean cloud temperature in the isothermal regime. 
We now need to estimate the density $\rho_{ad}$ at which the gas becomes adiabatic. 
This estimate has been done by several authors \citep{rees,low76,Masunaga99} and consist
in estimating the cooling and freefall time. 
To estimate $\rho_{ad}$, we must compare the 
cooling time and the freefall time \citep{Masunaga99}. To estimate the former we simply compute the 
ratio of the thermal energy contained in a Jeans mass, defined as a sphere of radius 
$\lambda_J/2 = \sqrt{ \pi} C_s / 2 / \sqrt{G \rho_{ad}}$ to the energy radiated per units
of time, ${\cal L}$
\begin{eqnarray}
\label{etherm}
E_{therm} = {M_J \over m_p} \times {3 \over 2} k_B T_0,
\end{eqnarray}
\begin{eqnarray}
\label{Lrad}
{\cal L} \simeq 4 \pi \left( { \lambda_J  \over 2} \right)^2 {4 ac T_0 ^3 \over 3 \kappa_R(T_0) \rho_{ad}} { T_0 \over \lambda_J / 2 }. 
\end{eqnarray}
This last expression is simply the radiative flux through a sphere of diameter $\lambda_J$ in which 
the gradient $\partial _r T$ has been replaced by $T / r$, $\kappa_R$ is the Rosseland opacity, 
$c$ is the speed of light and $a= 4 \sigma / c$, $\sigma$ being the Stefan-Boltzmann constant. 
The cooling time is simply estimated as
\begin{eqnarray}
\label{cool}
\tau _{cool} = {E_{therm}  \over {\cal L} },
\end{eqnarray}
while the freefall time is
\begin{eqnarray}
\label{ffall}
\tau _{ff} = \sqrt{ 3 \pi  \over 32 G \rho_{ad} }.
\end{eqnarray}
Requiring that $\tau_{cool} \simeq \tau_{ff}$, we get 
\begin{eqnarray}
\label{rhoad}
\rho_{ad} = \left( {32 \over 3 \pi} G m_p^4  k_B^{-4} a^2 c^2 \right)^{1/3} T_0^ {4/3} \kappa_R^{-2/3}. 
\end{eqnarray}

\subsection{Results, prediction and qualitative comparisons}
Combining Eqs.~(\ref{hydrostat2}),~(\ref{rho_T}),~(\ref{rhoad}) we get an expression for $M_L$
\begin{eqnarray}
\nonumber
M_L  &\simeq   {\cal H}^{3/2}  {5^{3/2} 3^{2/3} \over   2^{11/6} \pi^{1/3} }
\left( G^{10} m_p^{13}  k_B^{-13} a^2 c^2 \right)^{-1/6} 
 T_{dis}^{1/4}  T_{ex}^{1/2} 
 T_0^ {1/12} \kappa_R(T_0)^{1/3} \nonumber\\
&= {\cal H}^{3/2} \times 0.08 M_\odot \left({ T_0 \over 10 {\rm K}} \right)^{1/12}  \left( { \kappa_R(T_0) \over 0.1 {\rm cm}^2 \, {\rm g}^{-1}} \right)^{1/3}. 
\label{mass_lars}
\end{eqnarray}
 The estimated value of $M_L$ from numerical simulations is about 0.03 $M_\odot$ \citep{Vaytet17}, 
thus  ${\cal H} \simeq 0.5$ leads to the correct value. 
Indeed integrating the proper hydrostatic equilibrium (see Eq.~7 of paper II)
starting from $\rho_{dis}$ and using the piecewise polytropic equation of state discussed in the text, we obtain 
a mass equal to 0.03 $M_\odot$. The typical value of $\kappa_R$ is taken from \citet{semenov2003} at 10 K.
We recall that the interest of Eq.~(\ref{mass_lars}) is to explicit the various dependences.

Equation~(\ref{mass_lars}) makes two important predictions. First it does not depend significantly 
of $T_0$. Second it depends on the opacity $\kappa_R$, and therefore on the metallicity, $Z$, to the power 1/3,
which is rather shallow. This implies that one would not expect an important dependence 
of the peak of the IMF on the metallicity. Typically changing $Z$ by a factor 10 implies a shift of the $M_L$ of about 2 if 
$T_0$ is unchanged. There is however likely another effect due to the fact that at low metallicity the 
temperature $T_0$ is higher because the cooling is lower. Yet the opacity increases with the 
temperature. This effect is particularly visible in the simulations presented by \citet{bate2014} where 
four runs with four different metallicities are presented with $Z$ going from 0.01 to 3. The opacities 
are proportional to $Z$ making large variation of $\kappa_R$. However, the author found that the background 
temperature varies from 10 K to 70 K at $Z=0.01$. Looking at his figure 1, we see that the 
opacity at 10 K and $Z=1$ and the one at 70 K and $Z=0.01$,  are nearly identical and equal to 0.02-0.03 g cm$^{-2}$. 
According to the dependence stated by Eq.~(\ref{mass_lars}) and doing the proper hydrostatic calculation this leads 
to an estimate of $M_L \simeq 0.02~M_\odot$ and therefore to a peak of the IMF around 10 times this value (as argued in paper II and 
below). This is clearly not incompatible with Fig.~6 of \citet{bate2014} where we see that in all cases
there is a peak located at a few 0.1 $M_\odot$. 

\subsection{The cooling time of the first hydrostatic core}
A central aspect of the physics of the first hydrostatic core is that its lifetime is significantly 
longer than the dynamical time. Here we provide an estimate and a discussion on the implication. 

To get the cooling time, we again use the thermal energy and the radiative flux. However we now 
take the typical values  that correspond to the inner part of the core, say within 1 AU or so. 
\begin{eqnarray}
\label{etherm_L}
E_{therm} \simeq {M_L \over m_p} \times {5 \over 2} k_B T_{dis},
\end{eqnarray}
the factor 5 is due to the $H_2$ molecule being diatomic as in this regime rotational levels
are excited \citep[see e.g.][for a complete treatment]{saumon1995}. 
\begin{eqnarray}
\label{Lrad_L}
{\cal L} \simeq 4 \pi R_L ^2 {4 ac T_{dis} ^3 \over 3 \kappa_R(T_{dis}) \rho_{dis}} { T_{dis} \over R_L }. 
\end{eqnarray}

Using again $M_L = 4 \pi / 3 \rho R^3,  $ to infer the density in Eq.~(\ref{Lrad_L})
and Eq.~(\ref{hydrostat}) to infer $R T_{dis}$,
 we get
\begin{eqnarray}
\label{cool_t}
\tau_{cool} &=& {3^2   5^5 \over 2^7 \pi^2} {k_B^5 T_{dis} \over a c G^4 m_p^5} {\kappa_R (T_{dis}) \over M_L^2} \\
&\simeq& 5.6 \times 10^4 {\rm yr} {\kappa_R (T_{dis}) \over 10 \, {\rm g \, cm}^{-2} } \left( { M_L \over 0.03 \, {\rm M}_\odot  } \right) ^{-2}.
\nonumber
\end{eqnarray}
where $\kappa_R(T_{dis}) = 10$  cm$^{2}$ g$^{-1}$ is from \citet{semenov2003}.
By contrast the freefall time at a density of $\rho_{dis}$ is about 1 yr. 
More generally a freefall time of 5$\times$10$^4$ yr corresponds to a density of roughly 10$^6$ cm$^{-3}$, 
which is far below the density of the first hydrostatic core. 

The other relevant time is the accretion one, $\tau_{acc}$, that is to say the time necessary to accumulate a mass 
equal to $M_L$
\begin{eqnarray}
\label{accret_t}
\tau_{acc} = {M_L \over  {d M \over dt} } =  300 \, {\rm yr} {M_L \over 0.03 {\rm M} _\odot}  \left(  {{d M \over dt} \over 10^{-4} {\rm M} _\odot {yr}^{-1}} \right) ^{-1},
\end{eqnarray}
which is very close to the estimated life time of the Larson core \citep{Vaytet17} of 100-1000 yr since the typical 
accretion rate 
inferred in collapse calculations is a few $10^{-5} \; {\rm M} _\odot {yr}^{-1}$ \citep{foster1993,h2003,Vaytet17}.

This implies that the dynamics of the gas that enters this core is halted. The core stays a period of time 
that is long with respect to the dynamical timescale which itself is typical for a self-gravitating object of this 
density. Moreover the cooling time is even longer for an object with a smaller mass ($\tau_{cool} \propto M^{-2}$). 
Therefore although in practice objects with a mass smaller than $M_L$ could cool down and  eventually form 
a star or a brown dwarf, in practice it is very difficult if the object is not isolated. There is plenty of time 
for the object to accrete and acquire a mass of at least $M_L$ which would trigger H$_2$ dissociation 
and second collapse. 

 Equations~\ref{cool_t} and~\ref{accret_t} predict a change of behaviour of the lifetime of the first hydrostatic
 core at a mass $M \simeq M_L$. This transition is clearly seen in the recent work by \citet{stamer2018} who 
presented a series of 1D calculation (see their Figure~6). 

We stress in particular that the 
minimum mass for fragmentation \citep{rees,low76,Masunaga99}, 
estimated to be the Jeans mass at the density $\rho_ {ad}$ (as stated from Eq.~\ref{rhoad}),
is 
not clear because the fragments have typically a mass of about $10^{-3}$ M$_\odot$, which is too low to 
 induce the dissociation of H$_2$ and form a protostar. In most cases, these fragments will simply grow in mass
and form a first hydrostatic core or will merge with an existing one or a protostar. Only if the fragments are 
very isolated, they will have the opportunity to cool.

\section{Sink velocity as a function of age and its dependence to the number of neighbours}
\label{sink_vel}

\setlength{\unitlength}{1cm}
\begin{figure*}
\begin{picture} (0,19)
\put(0,12){\includegraphics[width=8.7cm]{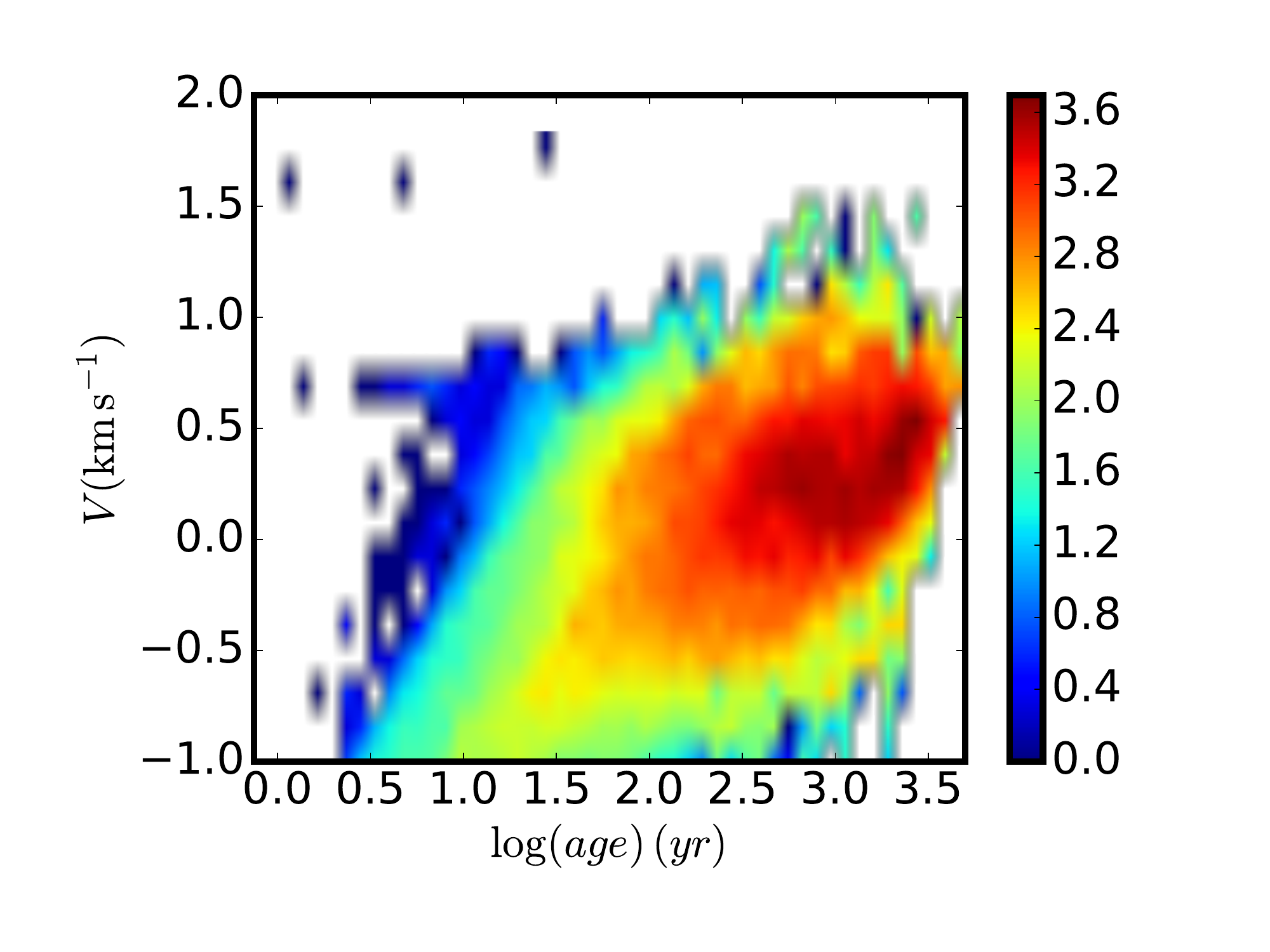}}  
\put(9,12){\includegraphics[width=8.7cm]{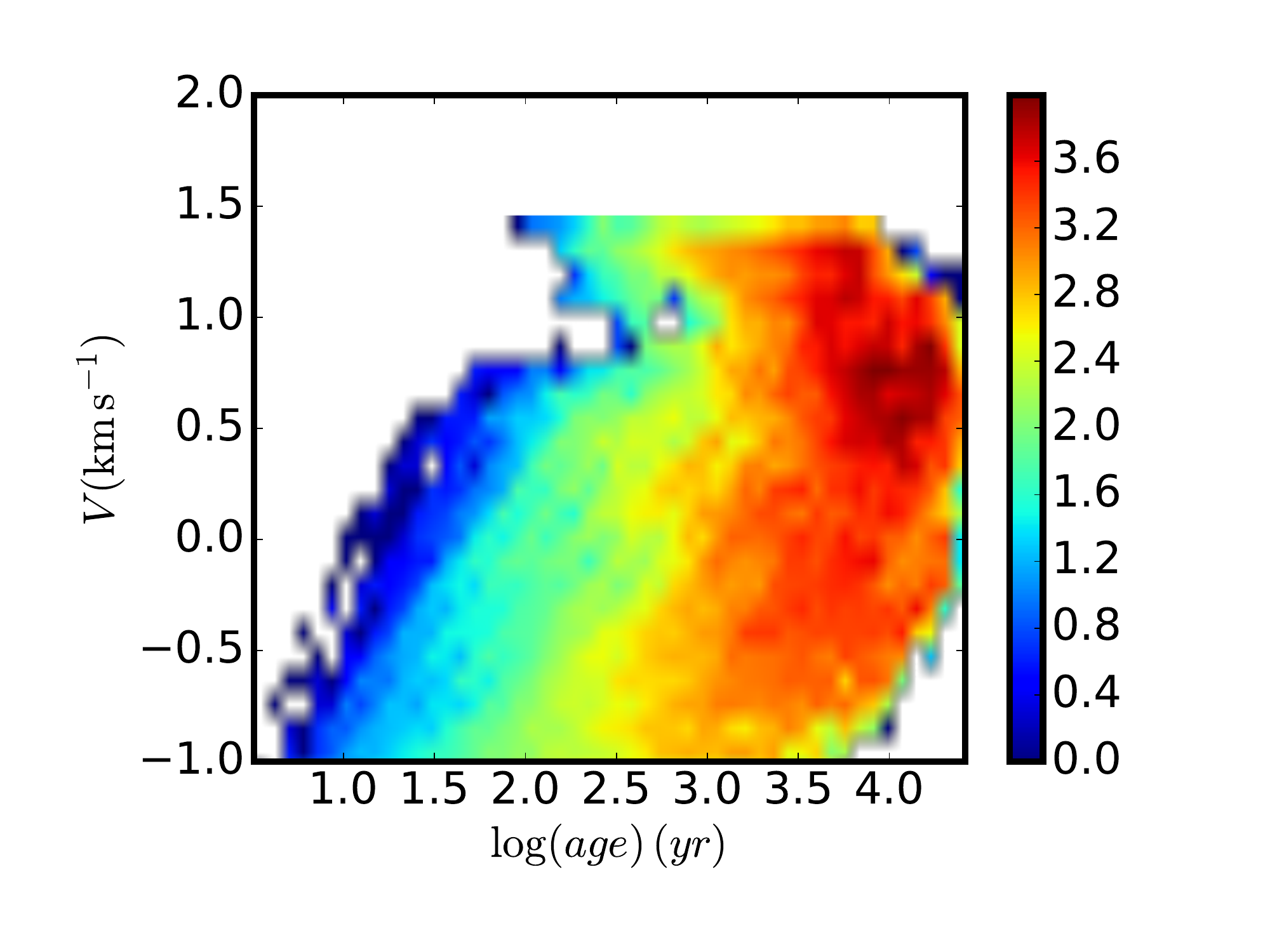}}  
\put(9,6){\includegraphics[width=8.7cm]{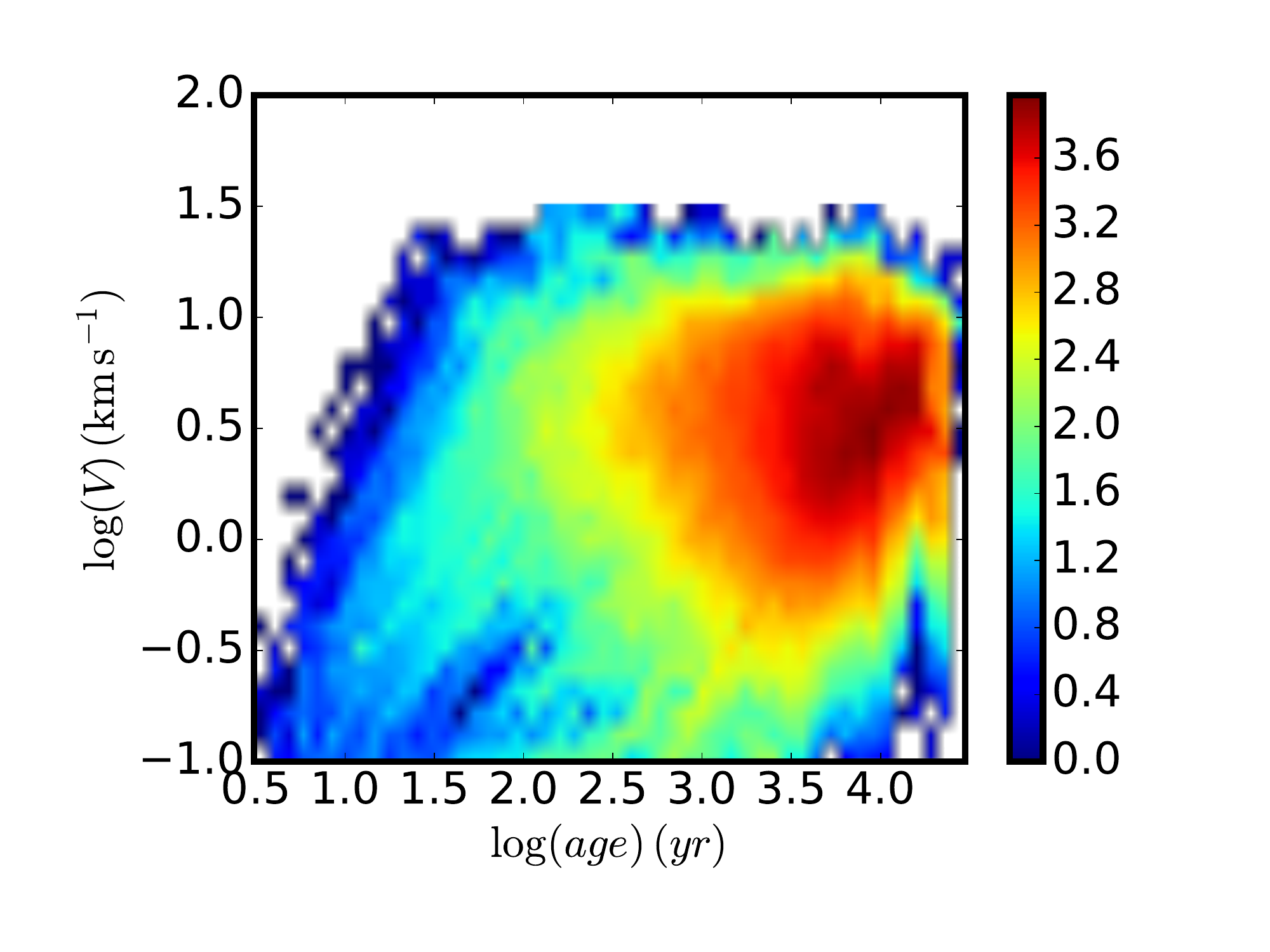}}  
\put(0,6){\includegraphics[width=8.7cm]{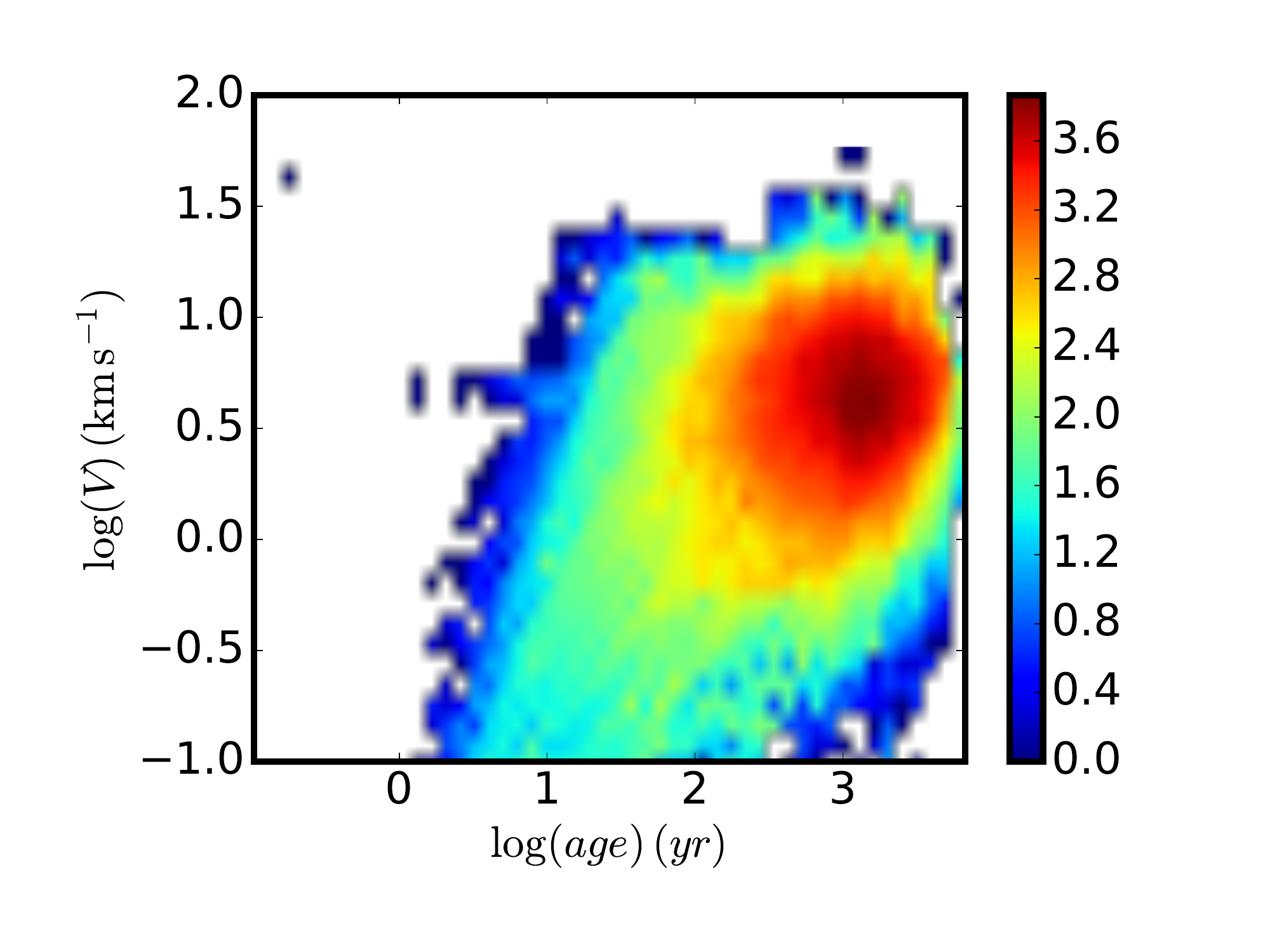}}  
\put(9,0){\includegraphics[width=8.7cm]{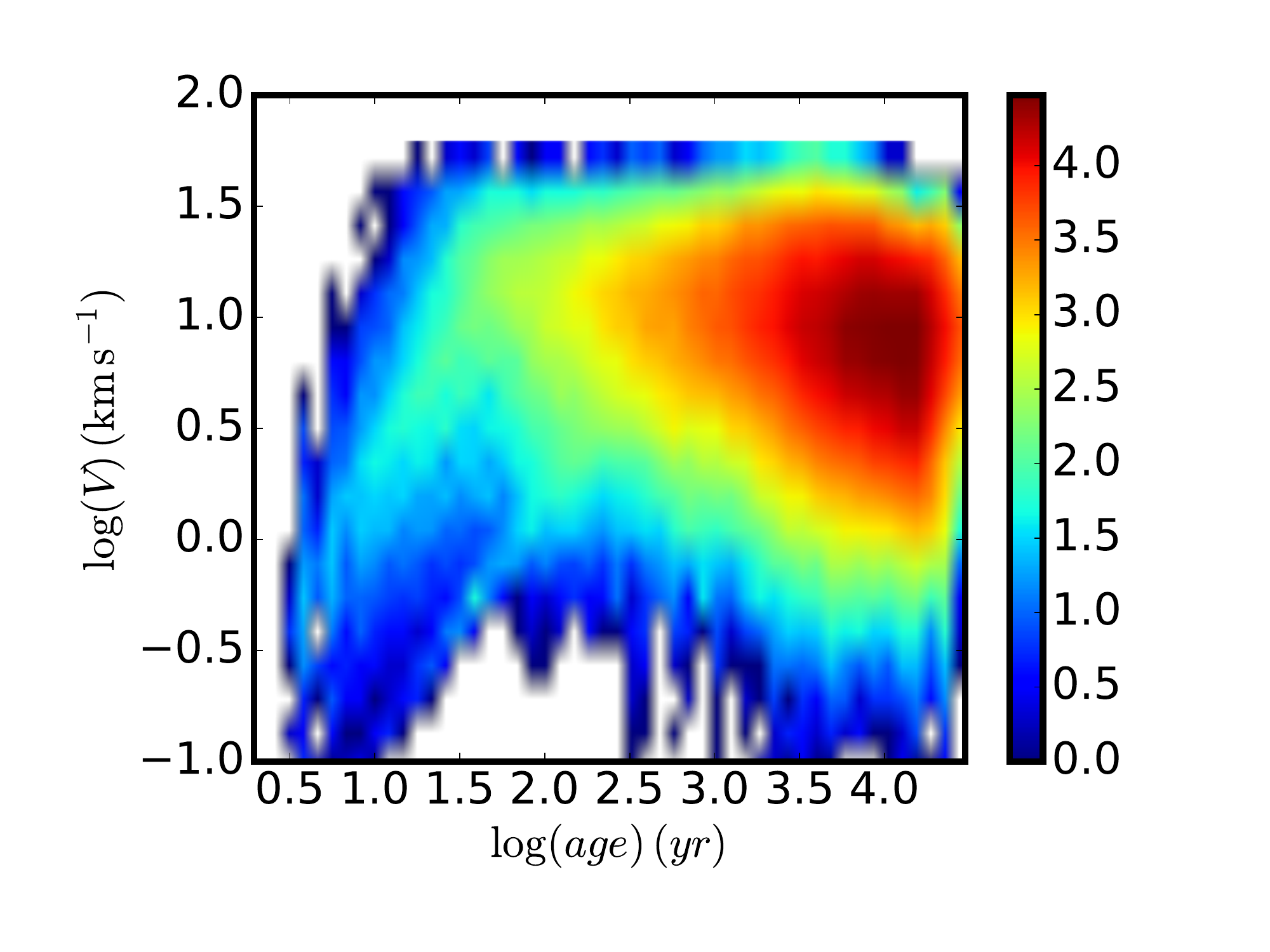}}  
\put(0,0){\includegraphics[width=8.7cm]{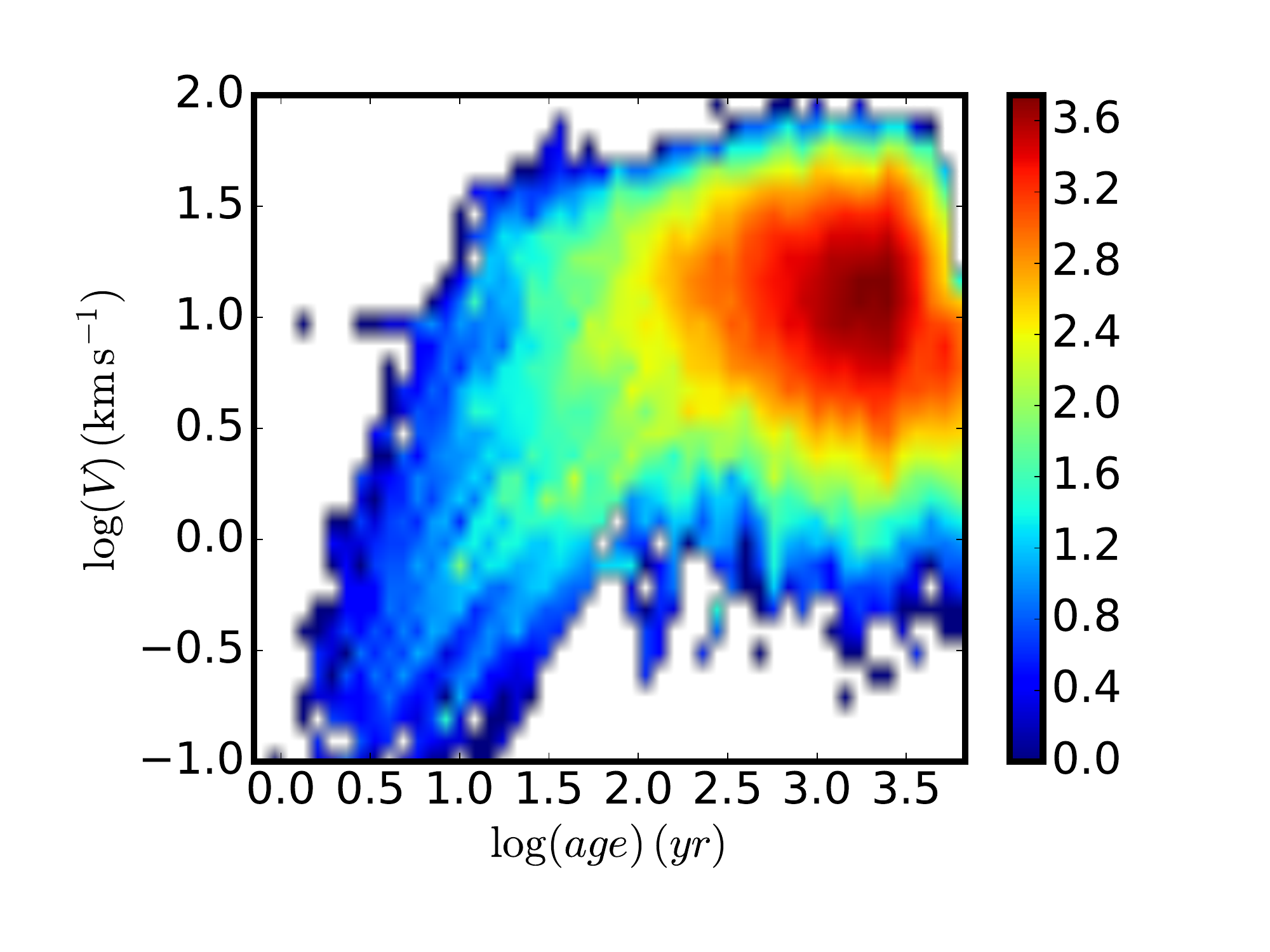}}  
\put(2,18){run 1, no neighbour}
\put(11,18){run 2, no neigbour}
\put(2,12){run 1, number of neighbours =2-4}
\put(11,12){run 2, number of neighbours =2-4}
\put(2,6){run 1, number of neighbours $>$ 4}
\put(11,6){run 2, number of neighbours $>$ 4}
\end{picture}
\caption{Bidimentional histogram of the sink velocity 
variation (i.e. with respect to the velocity at the sink birth) vs the sink age (i.e. time
since its birth). Top: sinks having no neighbour within 200 AU, middle: sinks having a number 
of neighbours between 2 and 4 and bottom: sinks having more than 4 neigbours within 200 AU. 
Left: run 1, right: run 2.}
\label{run1_vclump}
\end{figure*}
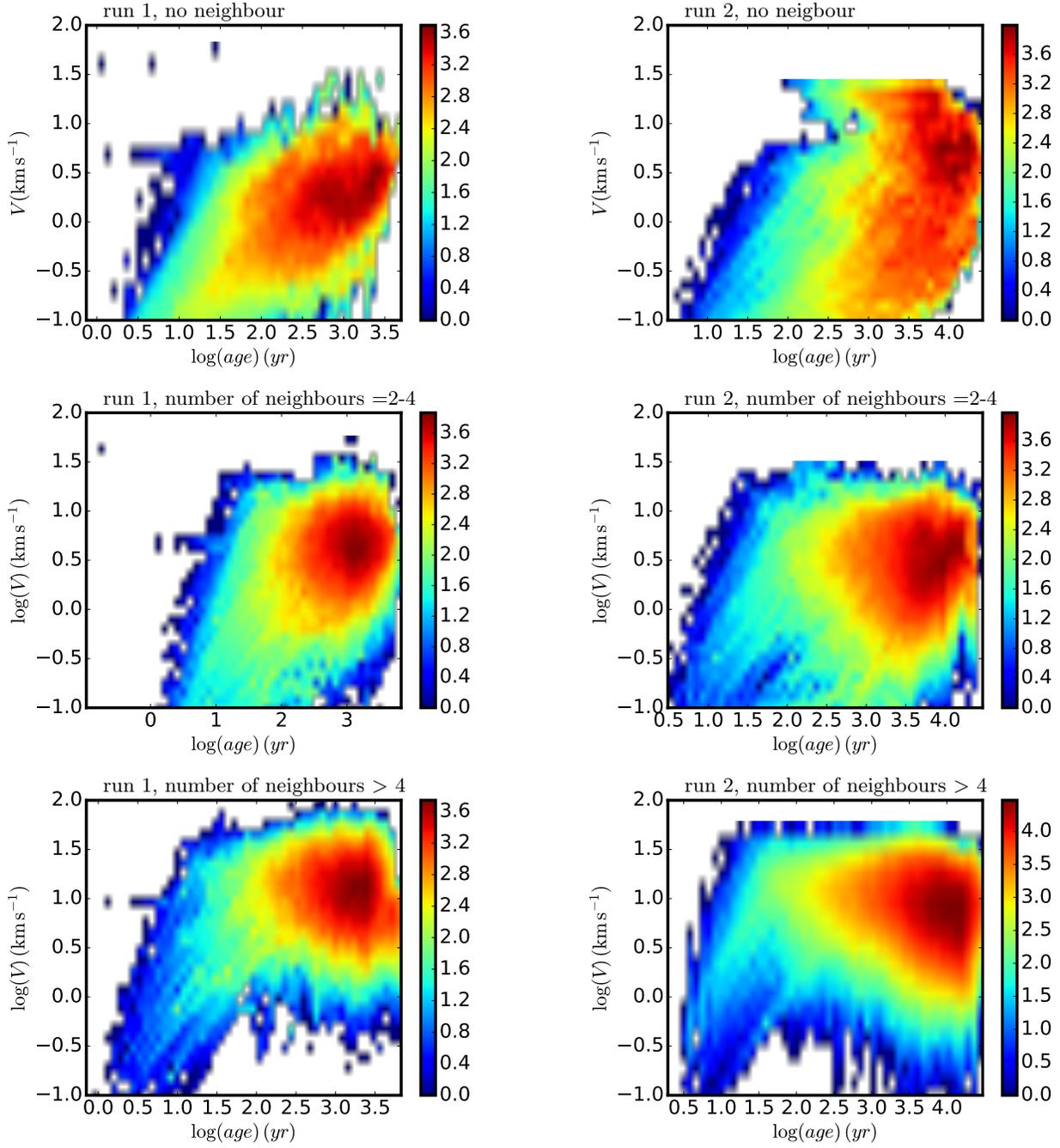

To investigate the dynamical effect that sink particle neighbours may have 
on each other, we have computed the velocity variation (i.e. the 
velocity with respect to the sink initial velocity) as a function 
of sink age. The idea is to determine how fast the velocity of the sink 
changes with respect to its initial environment and whether the effect depends
on the number of neighbours. 
Figure~\ref{run1_vclump} display bidimentional 
histograms for run 1 and run 2 and for the sinks having 0, 2-4 or 
more than 4 neighbours within 200 AU. 
Clearly we wee that the sink velocity variation increases
with time reaching values on the order of 3-10 km s$^{-1}$
in a few kyr. This time is comparable to the accretion time 
of the sink particles. We also see that the velocity variation 
is stronger when the sinks have more neighbours. This supports the idea 
that the neighbours limit the accretion into the sinks by two ways. 
First they screen the gas that would otherwise have fallen  into the primary 
sink and second they push it away from its initial gas reservoir.

\section{Estimating the external pressure in the virial theorem}
\label{vir_expr}
As discussed in section~\ref{insta_thres}, the turbulent pressure could have 
a very strong impact on the result by decreasing drastically the number 
of fragments. Therefore a more detailed analysis, particularly on the role of the external 
pressure that must also be taken into account, is required.

We remind the virial theorem 
\begin{eqnarray} \label{eq_vir_comp} 
{ 1 \over 2 } { \partial ^2 I \over \partial t^2} =  -E_\mathrm{g} + 2E_\mathrm{th} 
+ 2E_\mathrm{kin} - \oint P_{th} \, {\bf r . dS} -  \oint \rho \, ({\bf v . r})  ({\bf v . dS}),
\end{eqnarray} 
where the various term are the second derivative of the inertia momentum, 
the gravitational energy, the thermal and the kinetic ones while 
the two last ones  stand for the thermal and ram pressure in the outskirt of the {\it cores}.
We remind that the last term comes from the equality:
\begin{eqnarray}
\int \rho {d {\bf v} \over dt} . {\bf r} dV  = { 1 \over 2 } { \partial ^2 I \over \partial t^2}
- 2E_\mathrm{kin} + \int \nabla.( \rho ({\bf v . r})  {\bf v} ) dV. 
\end{eqnarray} 

Below, we seek for expressions of
$2E_\mathrm{ther}$, $2E_\mathrm{kin}$, ${\rm VP}_{ext}$   and ${\rm VP}_{ram}$. Assuming 
spherical symmetry, we have
\begin{eqnarray}
{\rm VP}_{ext} = \oint  P_{th} \, {\bf r . dS} = 3  P_{ext} V_{vol}, 
\end{eqnarray} 
where $V_{vol}$ is the volume and 
\begin{eqnarray}
{\rm VP}_{ram} =  \oint \rho \, ({\bf v . r})  ({\bf v . dS})  = 3  P_{ram} V_{vol}, 
\end{eqnarray} 
where $P_{ram}= \rho _{ext} v_n^2$ is the ram pressure at the clump boundary. 

Let us consider a spherical clump of external radius $r_0$, 
outer density $n_0$, temperature $T_0$ and of typical velocity dispersion $\sigma_0$. 
To estimate the various contribution we also need to know the density 
distribution within the clump that for the sake of simplicity we assume
to be $n(r) \propto r^{-\alpha}$, with $\alpha$ ranging typically between 0 and 2. 
We also assume that $\sigma(r) = \sigma_0 (r/r_0)^{0.5}$. 
 The external density is  $n_{ext}$. Typically we expect it to be related to the 
outer density (i.e. the density at the inner edge of the clump), $n_0$,  
by a factor on the order of ${\cal M}^2$, the typical Mach number at the clump scale.
We have $n_{ext}= {\cal C} n_0$, where ${\cal C}$  ranges between 0 and 1. Given the Mach 
numbers which at the scale of the clumps  are on the order of 1-2, this leads 
to a contrast,  ${\cal C}$, of $0.25-1$. This contrast however is along the direction 
of the confining shock and therefore for a spherical clump is less pronounced. 

This leads to 

\begin{eqnarray} \label{etherm} 
 2E_\mathrm{ther}  = {9 \over 3 - \alpha} M_0  C_{s,0}^2 ,  
\end{eqnarray}
where $M_0 = 4 \pi / 3  r_0^3 n_0 m_p$, $C_{s,0}^2= k_b T_0 / m_p$, 

\begin{eqnarray} \label{turb} 
 2E_\mathrm{kin}  = {3 {\cal M}^2 \over 4 - \alpha} M_0  C_{s,0}^2,  
\end{eqnarray} 

\begin{eqnarray} \label{pext_th} 
3 P_{ext} V  = 3 {\cal C} M_0  C_{s,0}^2,  
\end{eqnarray} 
and the ram pressure
\begin{eqnarray} \label{pext_ram} 
3 P_{ram} V_{vol}  =  \oint \rho_0 v_n^2 r ds  = 3 {\cal C} M_0  \sigma_{0,sph}^2 = 3 {\cal C} {\cal M}_{sph}^2  M_0  C_{s,0}^2 ,  
\end{eqnarray} 
where $\sigma_{0,sph}^2 = \oint  v_n^2 r ds / \oint r ds$ is the spherical part of the velocity field at the 
surface of the clump.

\begin{eqnarray} \label{val_vir} 
 2E_\mathrm{ther} + 2E_\mathrm{kin} - 3 P_{ext} V =  3  M_0  C_{s,0}^2 \times \left( 
{3 \over 3 - \alpha} - {\cal C}  + { {\cal M}^2 \over 4 - \alpha} -  {\cal C} {\cal M}_{sph}^2 \right).
\end{eqnarray}

Thus we see that for $\alpha=1$, the effective thermal energy, corrected from the external pressure 
should be multiplied by $3 / ( 3 - \alpha) - {\cal C} $ while the effective kinetic energy should be 
multiplied by  
$  1 -  3 {\cal C} {\cal M}_{sph}^2 / {\cal M}^2$.
The external pressure exerted by the velocity at the edge of the clump reduces the turbulent pressure.
 If the spherical mode represents one third of the 
energy and if ${\cal C}=1$, then the turbulent pressure cancels out exactly. The exact value of 
$1 - 3 {\cal C} {\cal M}_{sph}^2 / {\cal M}^2$ is hard to infer analytically but it seems likely that 
the value $2 E_{kin}$ overestimates the turbulent pressure. Let us stress in particular, that if there is no clear 
contrast between the external density and the clump density, i.e. ${\cal C} \simeq 1$, then the value of 
$1 - 3 {\cal C} {\cal M}_{sph}^2 / {\cal M}^2 \simeq 0$ since 
$ {\cal M}_{sph}^2 \simeq {\cal M}^2 / 3$ (as it is the case for a plane parallel converging flow for example).

\section{Envelope temperature}
\label{env_temp}
Strictly speaking,
the isothermal assumption is no longer valid due to
the heating induced by the gravitational contraction and the fact that the optical depth becomes 
close to or larger than 1. To address this issue we use the effective equation of state inferred to fit the thermal behaviour of dense material \citep{Masunaga98,vaytet2013,Vaytet17}, used in the present simulations  namely
\begin{eqnarray}
T = T_0 \left[1 + (\rho / \rho _{ad})^{(\gamma-1)}\right], \label{eq_eos}
\end{eqnarray}
where we adopt $T_0=10$ K, $\rho_ {ad}= 3.4 \times 10^ {-14}$ g cm$^{-3}$ and 
$\gamma = 5/3$. This expression holds at early time close to the 
formation of the Larson core.  At later time, the temperature increases further but here we are concerned
with the very early fragmentation stages for low mass stars, i.e. having masses of a few 0.1 $M_\odot$.
 Moreover, the increase of temperature remains limited for most of the dense gas.
When solving Eq.~(\ref{eq_int_vir}), with the condition that the 
mass of the fluctuation is $M_p = M_L$, we  find either 0 or 2 solutions, corresponding 
respectively to $\rho_{min}$ and $\rho_{max}$. This stems from the fact that at high density, the 
thermal pressure always dominates gravity due to the stiff value of $\gamma$. The cases without solution
arise when $A$ is high typically $> 20$. In practice, in the relevant regime, we found out that this does not yield
significant differences compared to the isothermal case. The reason is that the typical 
distance between the shielding fragments is 3-10 $r_{e,L}$, which corresponds to 
a distance of 60-200 AU. For $A=10$ this corresponds to densities $3 \times 10^8$-$3 \times 10^9$ cm$^{-3}$,
at which the gas is nearly isothermal.

\end{document}